\documentclass[12pt]{article}
\usepackage[latin9]{inputenc}
\usepackage{geometry}
\usepackage{float}
\usepackage{mathtools}
\usepackage{amsmath}
\usepackage{amsthm}
\usepackage{amssymb}
\usepackage{graphicx}
\usepackage{setspace}
\usepackage{esint}
\usepackage[authoryear]{natbib}
\usepackage{url} 
\usepackage{booktabs} 
\usepackage{subcaption}

\makeatletter

\providecommand{\tabularnewline}{\\}
\floatstyle{ruled}

\newfloat{algorithm}{tbp}{loa}
\providecommand{\algorithmname}{Algorithm}
\floatname{algorithm}{\protect\algorithmname}

\usepackage{latexsym}
\usepackage{amsthm}\usepackage{amsfonts}\usepackage{graphicx}\usepackage{epsfig}
\usepackage{bm}
\newcommand*{\patchAmsMathEnvironmentForLineno}[1]{%
      \expandafter\let\csname old#1\expandafter\endcsname\csname #1\endcsname
      \expandafter\let\csname oldend#1\expandafter\endcsname\csname end#1\endcsname
      \renewenvironment{#1}%
         {\linenomath\csname old#1\endcsname}%
         {\csname oldend#1\endcsname\endlinenomath}}%
    \newcommand*{\patchBothAmsMathEnvironmentsForLineno}[1]{%
      \patchAmsMathEnvironmentForLineno{#1}%
      \patchAmsMathEnvironmentForLineno{#1*}}%
    \AtBeginDocument{%
    \patchBothAmsMathEnvironmentsForLineno{equation}%
    \patchBothAmsMathEnvironmentsForLineno{align}%
    \patchBothAmsMathEnvironmentsForLineno{flalign}%
    \patchBothAmsMathEnvironmentsForLineno{alignat}%
    \patchBothAmsMathEnvironmentsForLineno{gather}%
    \patchBothAmsMathEnvironmentsForLineno{multline}%
    }
\usepackage[mathlines,displaymath]{lineno}

\include{def}
\def\dispmuskip{\thinmuskip= 3mu plus 0mu minus 2mu \medmuskip=  4mu plus 2mu minus 2mu \thickmuskip=5mu plus 5mu minus 2mu}
\def\textmuskip{\thinmuskip= 0mu                    \medmuskip=  1mu plus 1mu minus 1mu \thickmuskip=2mu plus 3mu minus 1mu}
\def\beq{\dispmuskip\begin{equation}}    \def\eeq{\end{equation}\textmuskip}
\def\beqn{\dispmuskip\begin{displaymath}}\def\eeqn{\end{displaymath}\textmuskip}
\def\bea{\dispmuskip\begin{eqnarray}}    \def\eea{\end{eqnarray}\textmuskip}
\def\bean{\dispmuskip\begin{eqnarray*}}  \def\eean{\end{eqnarray*}\textmuskip}

\newcommand{\wh}{\widehat}

\def\v{\boldsymbol}

\def\E{{\mathbb E}}                         
\def\V{{\mathbb V}}

\def\v{\boldsymbol}

\def\bs{\boldsymbol}

\def\thanks#1{\protected@xdef\@thanks{\@thanks
        \protect\footnotetext{#1}}}

\usepackage[mathlines,displaymath]{lineno}

\makeatother

\begin{document}

\title{Subsampling Sequential Monte Carlo for Static Bayesian Models}
\date{\empty}
\author{David Gunawan$ ^{1,4}$, Khue-Dung Dang$ ^{2, 4}$, Matias Quiroz$ ^{2,4,5}$, \\ Robert Kohn$ ^{3,4}$ and Minh-Ngoc Tran$ ^{4,6}$\thanks{~$^1$:\textit{School of Mathematics and Applied Statistics,University of Wollongong.} $^2$:\textit{School of Mathematical and Physical Sciences, University of Technology Sydney}  
$^3$:\textit{School of Economics, UNSW Business School,
University of New South Wales.} $^4$:\textit{ARC Centre of Excellence for Mathematical and Statistical Frontiers (ACEMS).} $^5$:\textit{Research Division, Sveriges Riksbank.} $^6$:\textit{Discipline of Business
Analytics, University of Sydney.} }}
\maketitle
\begin{abstract}
We show how to speed up Sequential Monte Carlo (SMC) for Bayesian inference in large data problems by data subsampling. SMC sequentially updates a cloud of particles through a sequence of distributions, beginning with a distribution that is easy to sample from such as the prior and ending with the posterior distribution. Each update of the particle cloud consists of three steps: reweighting, resampling, and moving. In the move step, each particle is moved using a Markov kernel; this is typically the most computationally expensive part, particularly when the dataset is large. It is crucial to have an efficient move step to ensure particle diversity.
Our article makes two important contributions. First, in order to speed up the SMC computation,
we use an approximately unbiased and efficient annealed likelihood estimator based on data subsampling. The subsampling approach is  more memory efficient than the corresponding full data SMC, which is an advantage for parallel computation.  %
Second, we use a Metropolis within Gibbs kernel with two conditional updates. A Hamiltonian Monte Carlo update makes distant moves for the model parameters, and a block pseudo-marginal proposal is used for the particles corresponding to the auxiliary variables for the data subsampling. We demonstrate both the usefulness and limitations of the methodology for estimating four generalized linear models and a generalized additive model with large datasets. 
\\\noindent \textbf{Keywords.}  Hamiltonian Monte Carlo, Large datasets, Likelihood annealing
\end{abstract}

\section{Introduction \label{sec:Introduction}}
The aim of Bayesian inference is to obtain the posterior distribution of unknown parameters, and in particular the posterior expectations of functions of the parameters. This is usually done by obtaining a simulation approximation of the expectation using samples from the posterior distribution. Exact approaches  such as Markov Chain Monte Carlo (MCMC) \citep{brooks2011handbook}
have been the main methods used for sampling from complex posterior distributions. Despite this, MCMC methods have some notable drawbacks and limitations. One drawback, often overlooked by practitioners when fitting complex models, is the failure to converge caused by poorly mixing chains. While Hamiltonian Monte Carlo \citep[HMC]{Neal:2011} is a remedy in many cases, it can be notoriously difficult to tune. Limitations of MCMC methods include the difficulties of assessing convergence, parallelizing the computation, and estimating the marginal likelihood efficiently from MCMC output,
the latter being useful for model selection \citep{Kass:1995}. Sequential Monte Carlo (see \citealp{doucet2001introduction} 
for an introductory overview) methods provide an alternative exact simulation approach to MCMC methods and overcome some of their drawbacks. Moreover, in contrast to MCMC methods, SMC can provide online updates of the parameters as data is collected, which is particularly useful for dynamic (time-varying parameters) models. SMC is also useful for static (non time-varying parameters) models \citep{chopin2002sequential, DelMoral:2006}, and can in such cases more easily explore multimodal posterior distributions than MCMC. Note that our definition of dynamic refers to the model parameters or any unobserved states being time-varying and not the data. For example, an autoregressive (AR) model is considered to be static as the parameters do not depend on time, whereas a state space model is considered to be dynamic since the states evolve through time.

Despite the advantages of SMC, it is remarkably less used than MCMC for static models. One possible explanation is that, while amenable to computer parallelization, it is still very computationally expensive and particularly so for large datasets. Another obstacle caused by large datasets is that they prevent efficient computer parallelization of SMC, as the full dataset needs to be available for each worker which is infeasible as it consumes too much Random-Access Memory (RAM). We propose an efficient data subsampling approach which significantly reduces both the computational cost of the algorithm and the memory requirements
when parallelizing: see Section \ref{subsec:MemoryEfficiencySubsamplingSMC} for a detailed explanation of the latter. Our approach utilizes the methods previously developed for Subsampling MCMC \citep{Quiroz2018, Dang2017} and places them within the SMC framework. See \cite{quiroz2018subsampling} for an introduction to Subsampling MCMC.

In the Bayesian context, SMC traverses a cloud of particles through a sequence of distributions, with the initial distribution both easy to sample from and to evaluate, while the final distribution is the posterior distribution. The cloud of particles at step $p$ is an estimate of the $p$th distribution in the sequence. The particles consist of the unknown parameters and any additional latent variables that are part of the model. The evolution of the particle cloud from one step to another consists of three steps: reweighting, resampling and moving. Of these, the first two steps are common to all SMC schemes and are straightforward. The move step is the most expensive and is critical to ensure that the particle cloud is representative of the distribution it aims to estimate.

To the best of our knowledge, data subsampling has not been explored in SMC. While \cite{wang2019annealed} term their algorithm Subsampling SMC, their approach is distinct as they combine data annealing and likelihood annealing, whereas we use data subsampling to estimate the likelihood. In particular, data annealing requires handling all the data, whereas the data subsampling approach only deals with a small fraction of the data at each stage. Specifically, we consider a likelihood annealing approach in which we estimate the annealed likelihood efficiently using an approximately unbiased estimator. Likelihood estimates for SMC in a non-subsampling context have been used in \citet{Duan2015}, who propose to estimate the likelihood unbiasedly using a particle filter in a time series state space model application. However, \citet{Duan2015} use a random walk MCMC kernel for the move step of the model parameters, which is inefficient in high dimensions and we now turn to this issue.

The literature has focused on accelerating SMC algorithms by designing efficient MCMC kernels for the move step to achieve efficient particle diversity. Efficiency here means the ability of the MCMC kernel to generate distant proposals which have a high probability of being accepted. The advantage of an efficient move step is that few iterations of the kernel are needed, which is computationally cheap. Various approaches exist to achieve this. For example, adaptive SMC adapts the tuning parameters of the kernel to improve its efficiency \citep{jasra2011inference, fearnhead2013adaptive, buchholz2018adaptive}. \cite{south2016sequential} use SMC with a flexible copula based independent proposal, while \cite{sim2012information} and \cite{ south2017efficient} use derivatives to construct efficient proposals through the Metropolis Adjusted Langevin Algorithm \citep[MALA]{roberts2002langevin}. It is now well-known that the MALA proposal is a special case of the more general proposal utilizing Hamiltonian dynamics proposed in \cite{duane1987hybrid} (see \cite{Neal:2011} and \cite{betancourt2017conceptual} for an introduction to HMC). Although \cite{south2017efficient} mention HMC in their introduction, they only consider MALA in their paper and show how neural networks can be applied to adaptively choose its tuning parameters. \cite{daviet2016inference} considers HMC proposals for particle diversity. However, HMC is very slow for very large datasets and therefore this approach does not scale well in the number of observations.

We propose data subsampling to achieve scalability in the number of observations and HMC Markov move steps to achieve particle diversity. Section \ref{subsec:MemoryEfficiencySubsamplingSMC} shows that data subsampling lowers the memory requirements of the algorithm, making it possible to parallelise the computing on very large datasets. Our framework combines that of \citet{Duan2015} for carrying out SMC with an estimated likelihood, \citet{Quiroz2018} for estimating the likelihood and controlling the error in the target density and \cite{Dang2017} for constructing efficient proposals for high-dimensional targets in a subsampling context.

The rest of the article is organized as follows. Section \ref{sec:SMC} reviews SMC for static models. Section \ref{sec:Methodology} outlines the methodology. Section \ref{sec:Evaluations} applies the methodology in a variety of settings for simulated data. Section \ref{sec:RealApplication} presents an application of our method in model selection for a real dataset. Section \ref{sec:Conclusions} concludes.

\section{Sequential Monte Carlo}\label{sec:SMC}
\subsection{SMC for static Bayesian models}\label{subsec:SMC_static_models}
Denote the observed data $\boldsymbol{y}=(\boldsymbol{y}_1^\top, \dots, \boldsymbol{y}_n^\top)^\top$, with $\bs y_k  \in \mathcal{Y} \subset \mathbb{R}^{d_{\bs y}}$, where $\mathbb{R}^m$ is an $m$ dimensional Euclidean space. 
Let $\boldsymbol{\theta}$ be the vector of unknown parameters, $\bs \theta  \in 
\boldsymbol{\Theta} \subset \mathbb{R}^{d_{\boldsymbol{\theta}}}$, 
with $p(\boldsymbol{\theta})$ and $p\left(\boldsymbol{y}|\boldsymbol{\theta}\right)$ the prior and likelihood. In Bayesian inference, the uncertainty about $\boldsymbol{\theta}$ is specified by the posterior density 
$\pi(\boldsymbol{\theta})$, which by Bayes' theorem is
\begin{equation}\label{eq:posterior_density}
\pi(\boldsymbol{\theta}) =  \frac{p(\boldsymbol{\theta})p\left(\boldsymbol{y}|\boldsymbol{\theta}\right)}{p(\boldsymbol{y})},
\end{equation}
where $p\left(\boldsymbol{y}\right)=\int_{\boldsymbol{\Theta}}p\left(\boldsymbol{y}|\boldsymbol{\theta}\right)p\left(\boldsymbol{\theta}\right)d\boldsymbol{\theta}$ is the marginal likelihood which is often used for Bayesian model selection.

An important problem in Bayesian inference is to estimate the posterior expectation of a function $\varphi$ of $\boldsymbol{\theta}$,
\begin{equation}\label{eq:expectation_posterior_functional}
\E_{\pi}\left(\varphi(\boldsymbol{\theta})\right)=\int_{\Theta}\varphi\left(\boldsymbol{\theta}\right)\pi\left(\boldsymbol{\theta}\right)d\boldsymbol{\theta}.
\end{equation}
In simulation based inference, this is typically achieved by sampling from   \eqref{eq:posterior_density} and computing \eqref{eq:expectation_posterior_functional} by Monte Carlo integration. Another important problem is to compute the marginal likelihood in \eqref{eq:posterior_density}. However, it is well known that standard Monte Carlo integration is very inefficient for this task.

SMC \citep{doucet2001introduction, DelMoral:2006} is a collection of methods that provide a convenient approach to computing the posterior distribution and in addition the marginal likelihood. 
Likelihood tempered SMC specifies a sequence of $P$ densities, connecting the density of the prior $p(\boldsymbol{\theta})$ to the density of the posterior $\pi(\boldsymbol{\theta})$ in \eqref{eq:posterior_density}. The sequence is obtained through temperature annealing \citep{Neal:2001}, in which the likelihood is tempered as $p\left(\boldsymbol{y}|\boldsymbol{\theta}\right)^{a_p}$ with $a_0 = 0 < a_1 < \dots < a_P=1$. We note that frequently $P$ as well as $a_1, \dots, a_P$ are chosen adaptively as the SMC proceeds, and we do so in our article; see Section \ref{subsec:SMC_efficiency}. Our article estimates the tempered likelihood $p\left(\boldsymbol{y}
|\boldsymbol{\theta}\right)^{a_p}$ by data subsampling as in Section \ref{sec:Methodology}. The $p$th tempered posterior is
\begin{equation}\label{eq:tempered target density}
\pi_p(\boldsymbol{\theta}) = \frac{\eta_p(\boldsymbol{\theta})}{Z_p},\text{ where } \eta_p(\boldsymbol{\theta}) =  p\left(\boldsymbol{y}|\boldsymbol{\theta}\right)^{a_p}p(\boldsymbol{\theta})
\quad \text{and} \quad 
Z_p = \int_{\boldsymbol{\Theta}} p\left(\boldsymbol{y}|\boldsymbol{\theta}\right)^{a_p}p(\boldsymbol{\theta}) d\boldsymbol{\theta}.
\end{equation}

SMC starts by sampling a set of $M$ particles from the prior $p(\boldsymbol{\theta})$ and traverses them through the sequence of densities $\pi_p({\boldsymbol{\theta}}), p=1, \dots, P$ such that, for each $p$,
the reweighting, resampling and move steps are performed on the particles. Here, we assume for simplicity that  it is possible to sample from the prior; otherwise one can sample from some initial distribution $\pi_0({\boldsymbol{\theta}})$ whose support covers that of the prior $p({\boldsymbol{\theta}})$. At the final $p = P$, the particles are a (weighted) sample from $\pi(\boldsymbol{\theta})$. We now discuss this  in more detail.

The initial particle cloud and weights $\left\{ \boldsymbol{\theta}_{1:M}^{\left(0\right)},W_{1:M}^{\left(0\right)}\right\} $
are obtained by generating the $\left\{ \boldsymbol{\theta}_{1:M}^{\left(0\right)}\right\} $
from $p\left(\boldsymbol{\theta}\right)$, and giving them equal weight,
i.e., $W_{1:M}^{\left(0\right)}=1/M$. The weighted particles $\left\{ \boldsymbol{\theta}_{1:M}^{\left(p-1\right)},W_{1:M}^{\left(p-1\right)}\right\} $
at the $\left(p-1\right)$st stage, $p=1, \dots, P$, are (weighted) samples from $\pi_{{p-1}} \left(\boldsymbol{\theta}\right)$. 
At the $p$th stage, the transition from $\pi_{{p-1}}\left(\boldsymbol{\theta}\right)$ to $ 
\pi_{{p}}\left(\boldsymbol{\theta}\right)$ is obtained by the \textit{reweighting step}, 
$$w_i^{(p)} = W^{(p-1)}_i\frac{\eta_p\left(\boldsymbol{\theta}_i^{(p-1)}\right)}{\eta_{p-1}
\left(\boldsymbol{\theta}_i^{(p-1)}\right)} = W^{(p-1)}_i p\left(\boldsymbol{y}|\boldsymbol{\theta}_i^{(p-1)}\right)^{a_p-a_{p-1}}, $$
and then normalizing $W_i^{(p)} = w_i^{(p)}/\sum_{i^\prime=1}^M w_{i^\prime}^{(p)}$.
The reweighting
assigns vanishingly small weights to particles which are unlikely under the tempered likelihood. This might cause the so-called particle degeneracy problem, in which the weight mass is concentrated only on a small fraction of the particles, causing a small effective sample size (explained in Section \ref{subsec:SMC_efficiency}). This is resolved by the \textit{resampling step}, in which the particles $ \boldsymbol{\theta}_{1:M}^{\left(p\right)}$ are sampled with a probability equal to their normalized weights $W_{1:M}^{\left(p\right)}$, and then setting $W_{1:M}^{\left(p\right)}=1/M$. We use multinomial resampling for all the experiments and applications in the paper. While this ensures that the particles with small weights are eliminated, it causes the so-called particle depletion problem because resampling might lead to only a few distinct particles. This is resolved by the \textit{move step}, in which a $\pi_p$-invariant Markov kernel $K_p$ is applied to move each of the particles $R$ steps. Since a particle after the resampling step at stage $p$ is approximately a sample from $\pi_p(\theta)$ and $K_p$ is $\pi_p$-invariant, no burn-in period is required as in MCMC methods, where often a very large number of burn-in iterations are required. Finally, we note that the algorithm is easy to parallelize with respect to the $M$ particles, because the computations required for each particle do not depend on those of the other particles. Thus, provided that $p(\boldsymbol{y}|\boldsymbol{\theta})$ can be computed at each worker without storage issues, it is straightforward to implement the parallel version.

\citet{DelMoral:2006} provide consistency results and central limit theorems for estimating \eqref{eq:expectation_posterior_functional} based on the SMC output.

\subsection{Statistical efficiency of SMC}\label{subsec:SMC_efficiency}
The statistical efficiency of the $p$th stage of the SMC reweighting part is measured through the Effective Sample Size (ESS) defined as \citep{Liu:2001a}
$$\textrm{ESS}_p :=\left(\sum_{i=1}^{M}\left(W_{i}^{\left(p\right)}\right)^{2}\right)^{-1}.$$
The $\textrm{ESS}_p$ varies between $1$ and $M$, where a low value of $\textrm{ESS}_p$ indicates that
the weights are concentrated only on a few particles. It is necessary to choose the tempering sequence $\{a_p,p=1,\dots,P\}$ carefully because it has a substantial impact on the $\textrm{ESS}_p$. We follow \citet{DelMoral2012} and choose the tempering sequence
adaptively to ensure a sufficient level of particle diversity by selecting
the next value of $a_{p}$ such that 
$\textrm{ESS}_p$ stays close to some target value $\mathrm{ESS}_{\mathrm{target}}$; 
this is done by evaluating the $\textrm{ESS}_p$ over a grid points $a_{1:S,p}$ of potential values of $a_{p}$ for a given $p$ and selecting $a_{p}$ as  that value of $a_{s,p}$, $s = 1, \dots, S,$ whose $\textrm{ESS}_p$ is closest to $\mathrm{ESS}_{\mathrm{target}}$. Throughout our article $\mathrm{ESS}_{\mathrm{target}}=0.8M$.

For this adaptive choice of tempering sequence, \citet{beskos2016convergence} establish consistency results and central limit theorems for estimating  \eqref{eq:expectation_posterior_functional} based on the SMC output. Other adaptive methods to choose the tempering sequence such as the approach by \cite{DelMoral2012} may also be used instead.
\subsection{SMC estimation of the marginal likelihood}\label{subsec:marginal_likelihood_SMC}
The marginal likelihood $p\left(\boldsymbol{y}\right)$  is often used in the Bayesian literature to compare models by their posterior model probabilities \citep{Kass:1995}. An advantage of SMC is that it automatically produces an estimate of $p\left(\boldsymbol{y}\right)$. 

Using the notation of Section \ref{subsec:SMC_static_models}, $Z_{P} = p\left(\boldsymbol{y}\right)$, $Z_{0}=1$, and 
\[
p\left(\boldsymbol{y}\right)=\prod_{p=1}^{P}\frac{Z_{p}}{Z_{{p-1}}}\;\;\textrm{with}\;\;\frac{Z_{p}}{Z_{{p-1}}}=\int\left(\frac{\eta_{{p}}\left(\boldsymbol{\theta}\right)}{\eta_{{p-1}}\left(\boldsymbol{\theta}\right)}\right)\pi_{{p-1}}\left(\boldsymbol{\theta}\right)d\boldsymbol{\theta}.
\]
Because the particle cloud $\left\{ \boldsymbol{\theta}_{1:M}^{\left(p-1\right)}, W_{1:M}^{\left(p-1\right)}\right\}$ at the $(p-1)$st stage is an approximate sample from $\pi_{{p-1}}\left(\boldsymbol{\theta}\right)$,
the ratios above are estimated by
\[
\widehat{\frac{Z_{{p}}}{Z_{{p-1}}}}=\sum_{i=1}^{M}W_{i}^{\left(p-1\right)}\frac{\eta_{{p}}\left(\boldsymbol{\theta}_{i}^{\left(p-1\right)}\right)}{\eta_{{p-1}}\left(\boldsymbol{\theta}_{i}^{\left(p-1\right)}\right)},
\]
giving the estimate of the marginal likelihood
\begin{equation}\label{eq:marginal_likelihood_estimate}
\widehat{p}\left(\boldsymbol{y}\right)=\prod_{p=1}^{P}\widehat{\frac{Z_{p}}{Z_{p-1}}}.
\end{equation}

\section{Methodology \label{sec:Methodology}}
\subsection{Sequence of target densities}\label{subsec:target_densities}
Suppose that $\boldsymbol{y}_{k},k=1,...,n,$ are independent given $\boldsymbol{\theta}$ so that the likelihood and log-likelihood can be written as
\begin{equation}
L\left(\boldsymbol{\theta}\right)=\prod_{k=1}^{n}p\left(\boldsymbol{y}_{k}|\boldsymbol{\theta}\right) \text{ and } \ell\left(\boldsymbol{\theta}\right)=\sum_{k=1}^{n}\ell_{k}\left(\boldsymbol{\theta}\right),\label{eq:likelihood}
\end{equation}
where $\ell_{k}\left(\boldsymbol{\theta}\right)=\log p\left(\boldsymbol{y}_{k}|\boldsymbol{\theta}\right)$. We are concerned with the case where the log-likelihood is computationally
very costly, because $n$ is so large that repeatedly computing this
sum is impractical, or $n$ is moderately large but each term is expensive to evaluate. 

\cite{Quiroz2018} propose to subsample $m$ observations and estimate $L(\bs \theta)$ from an unbiased estimator  $\widehat{\ell}_m(\boldsymbol{\theta})$ of $\ell(\bs \theta)$ 
\begin{equation}
\widehat{L}\left(\boldsymbol{\theta}\right)= \exp\left(\widehat{\ell}_m(\boldsymbol{\theta}) - \frac{1}{2}\widehat{\sigma}_m^2(\boldsymbol{\theta}) \right),\label{eq:likelihood_estimator}
\end{equation}
where $\widehat{\sigma}_m^2(\boldsymbol{\theta})$ is an estimate of ${\sigma}^2(\boldsymbol{\theta}) = \V\left(\widehat{\ell}_m(\boldsymbol{\theta})\right)$. The motivation for \eqref{eq:likelihood_estimator} is that $\exp\left(\widehat{\ell}_m(\boldsymbol{\theta}) -\sigma^2(\boldsymbol{\theta})/2 \right)$ is an unbiased estimator of $L(\theta)$ when $\widehat{\ell}_m(\boldsymbol{\theta})$ is normal \citep{ceperley1999penalty}. We note that by the central limit theorem, $\widehat{\ell}_m(\boldsymbol{\theta})$ is likely to be normal for moderate $m$ when $n$ is large even if $m$ is a small fraction of $n$. More generally, \eqref{eq:likelihood_estimator} is an unbiased estimator for $L_{(m,n)}(\boldsymbol{\theta}):= \mathbb{E}\left(\widehat{L}\left(\boldsymbol{\theta}\right)\right)$, which we call the perturbed likelihood. The expectation with respect to the subsampling indices $\boldsymbol{u}$ is discussed below. \cite{Quiroz2018} show that when using the control variate in Section \ref{subsec:EfficientEstimator} in the estimator $\widehat{\ell}_m(\boldsymbol{\theta})$, and under some extra plausible assumptions, the fractional error of the perturbed likelihood is $$\left| \frac{L_{(m, n)}(\boldsymbol{\theta})-L(\boldsymbol{\theta})}{L(\boldsymbol{\theta})}\right| = O\left(\frac{1}{nm^2}\right).$$

Our approach is based on extending the target at the $p$th density, i.e. $\pi_p(\boldsymbol{\theta})$ in \eqref{eq:tempered target density}, to include the set of subsampling indices $\boldsymbol{u} = (u_1, \dots, u_m)$, where $\boldsymbol{u} \in \mathcal{U} \subset \{1, \dots, n\}^m$  when sampling data observations with replacement. Let $\widehat{L}_p(\boldsymbol{\theta})$ be an estimator of the tempered likelihood $L(\boldsymbol{\theta})^{a_p}$. Similarly to \cite{Quiroz2018}, we can unbiasedly estimate $a_p\ell(\boldsymbol{\theta})$ with $a_p\widehat{\ell}(\boldsymbol{\theta})$, and since $\mathbb{V}\left(a_p\widehat{\ell}(\boldsymbol{\theta})\right)=a^2_p \sigma^2(\boldsymbol{\theta})$ and motivated by \eqref{eq:likelihood_estimator}, we propose the annealed likelihood estimator  
\begin{equation}
\widehat{L}_p\left(\boldsymbol{\theta}\right)= \exp\left(a_p\widehat{\ell}_m(\boldsymbol{\theta}) - \frac{1}{2}a_p^2\widehat{\sigma}_m^2(\boldsymbol{\theta}) \right).\label{eq:tempered_likelihood_estimator}
\end{equation}
The extended target at the $p$th density is
\begin{equation}\label{eq:extended_target_at_sequence_p}
\overline{\pi}_{p}\left(\boldsymbol{\theta},\boldsymbol{u}\right)\propto \widehat{L}_p(\boldsymbol{\theta}) p\left(\boldsymbol{\theta}\right)p\left(\boldsymbol{u}\right) = \exp\left(a_p\widehat{\ell}_m(\boldsymbol{\theta}) - \frac{1}{2}a_p^2\widehat{\sigma}_m^2(\boldsymbol{\theta}) \right)
p\left(\boldsymbol{\theta}\right)p\left(\boldsymbol{u}\right),
\end{equation}
where $p(\boldsymbol{u})$ is the density of $\boldsymbol{u}$ (or, more correctly, a probability mass function since $\boldsymbol{u}$ is discrete). At the final annealing step, \eqref{eq:extended_target_at_sequence_p} becomes $\overline{\pi}_P\left(\boldsymbol{\theta},\boldsymbol{u}\right) \propto \widehat{L}\left(\boldsymbol{\theta}\right)p(\boldsymbol{\theta})p(\boldsymbol{u})$, which is the target considered in \cite{Quiroz2018}. \cite{Quiroz2018} show that the perturbed marginal density for $\boldsymbol{\theta}$, $\pi_{(m, n)}(\boldsymbol{\theta}) = \int_{\boldsymbol{\mathcal{U}}} \overline{\pi}_P\left(\boldsymbol{\theta},\boldsymbol{u}\right)d\boldsymbol{u}$ converges in the total variation metric to $\pi(\boldsymbol{\theta})$ at the rate $O\left(1/(nm^2)\right)$. Hence, our proposed approach is approximate but can be very accurate, while also scaling well with respect to the subsample size. For example, if we take $m = O(\sqrt{n})$, then by \citet[Part (i) of Theorem 1]{Quiroz2018}
\begin{equation*}
\int_{\boldsymbol{\Theta}} \left| \pi_{(m, n)}(\boldsymbol{\theta}) - \pi(\boldsymbol{\theta})\right| d\boldsymbol{\theta} = O\left(\frac{1}{n^2}\right).
\end{equation*}
Moreover, suppose that $\varphi(\boldsymbol{\theta})$ is a scalar function with finite second moment. Then, by \citet[Part (ii) of Theorem 1]{Quiroz2018}
\begin{equation*}
\left|\E_{\pi_{(m, n)}}\left(\varphi(\boldsymbol{\theta})\right) - \E_{\pi}\left(\varphi(\boldsymbol{\theta})\right)\right| = O\left(\frac{1}{n^2}\right).
\end{equation*}
Thus, the approximation obtained by our approach converges to the posterior (in total variation norm) at a very fast rate as do the posterior moment estimates. Sections \ref{sec:Evaluations} and \ref{sec:RealApplication} confirm empirically that we obtain very accurate estimates in most of our applications, even for an $m$ very small relative to $n$.

\subsection{Efficient estimator of the log-likelihood}\label{subsec:EfficientEstimator}
\cite{Quiroz2018} propose estimating $\ell(\boldsymbol{\theta})$ in \eqref{eq:likelihood} by the unbiased difference estimator,
\begin{equation}
\widehat{\ell}_{m}(\boldsymbol{\theta})=\sum_{k=1}^{n} q_k(\boldsymbol{\theta}) + \frac{n}{m}\sum_{j=1}^m \ell_{u_j}(\boldsymbol{\theta}) - q_{u_j}(\boldsymbol{\theta}), \quad u_j \in \{1, \dots, n\} \text{ iid}, \label{eq:difference_estimator_logl}
\end{equation}
where
$$\Pr(u_j = k) = \frac{1}{n}  \text{ for all } k = 1, \dots, n \text{ and } j=1, \dots,m,$$
and $q_{k}(\theta)\approx \ell_k(\theta)$ are control variates. The estimator is based on writing
\[
\ell\left(\boldsymbol{\theta}\right)=\sum_{k=1}^{n}q_{k}\left(\boldsymbol{\theta}\right)+\sum_{k=1}^{n}d_{k}\left(\boldsymbol{\theta}\right)=q\left(\boldsymbol{\theta}\right)+d\left(\boldsymbol{\theta}\right),
\]
with $d_{k}\left(\boldsymbol{\theta}\right)=\ell_{k}\left(\boldsymbol{\theta}\right)-q_{k}\left(\boldsymbol{\theta}\right)$,
$q\left(\boldsymbol{\theta}\right)=\sum_{k}q_{k}\left(\boldsymbol{\theta}\right)$,
and $d\left(\boldsymbol{\theta}\right)=\sum_{k}d_{k}\left(\boldsymbol{\theta}\right)$. The last term on the right hand side of \eqref{eq:difference_estimator_logl} is an unbiased estimator of $d(\boldsymbol{\theta})$. We now discuss a choice of control variates due to \cite{bardenet2017markov}, which computes $q(\boldsymbol{\theta})$ in $O(1)$ time. Hence, the cost of computing the estimator is $O(m)$ and we can take $m = O(\sqrt{n})$ in order to achieve the convergence rates $O(1/n^2)$ for both the perturbed density and its moments as discussed in Section \ref{subsec:target_densities}.

Let $\overline{\boldsymbol{\theta}}$ be an estimate of posterior location, for example the posterior mean, obtained from a current particle cloud from $\overline{\pi}_{{p}}\left(\boldsymbol{\theta},\boldsymbol{u}\right)$.
A second order Taylor series expansion of the log-density around $\overline{\boldsymbol{\theta}}$ is
\[
\ell_{k}\left(\boldsymbol{\theta}\right)=\ell_{k}\left(\overline{\boldsymbol{\theta}}\right)+\nabla_{\boldsymbol{\theta}}\ell_{k}\left(\overline{\boldsymbol{\theta}}\right)^{\top}
\left(\boldsymbol{\theta}-\overline{\boldsymbol{\theta}}\right)+\frac{1}{2}\left(\boldsymbol{\theta}-\overline{\boldsymbol{\theta}}\right)^{\top}\left(\nabla_{\boldsymbol{\theta}\boldsymbol{\theta}^{\top}}^{2}\ell_{k}\left(\overline{\boldsymbol{\theta}}\right)\right)\left(\boldsymbol{\theta}-\overline{\boldsymbol{\theta}}\right)+o\left(||\boldsymbol{\theta}-\overline{\boldsymbol{\theta}}||\right),
\]
where $o\left(\delta\right)$ means that $o\left(\delta\right)/\delta\rightarrow0$ as $\delta\rightarrow0$. We approximate $\ell_k(\boldsymbol{\theta})$ by
\[
q_{k}\left(\boldsymbol{\theta}\right)=\ell_{k}\left(\overline{\boldsymbol{\theta}}\right)+\nabla_{\boldsymbol{\theta}}\ell_{k}\left(\overline{\boldsymbol{\theta}}\right)^{\top}\left(\boldsymbol{\theta}-\overline{\boldsymbol{\theta}}\right)+\frac{1}{2}\left(\boldsymbol{\theta}-\overline{\boldsymbol{\theta}}\right)^{\top}\left(\nabla_{\boldsymbol{\theta}\boldsymbol{\theta}^{\top}}^{2}\ell_{k}\left(\overline{\boldsymbol{\theta}}\right)\right)\left(\boldsymbol{\theta}-\overline{\boldsymbol{\theta}}\right).
\]
Then,
\begin{eqnarray*}
q\left(\boldsymbol{\theta}\right)  & = & A\left(\overline{\boldsymbol{\theta}}\right)+B\left(\overline{\boldsymbol{\theta}}\right)\left(\boldsymbol{\theta}-\overline{\boldsymbol{\theta}}\right)+\frac{1}{2}\left(\boldsymbol{\theta}-\overline{\boldsymbol{\theta}}\right)^{\top}C\left(\overline{\boldsymbol{\theta}}\right)\left(\boldsymbol{\theta}-\overline{\boldsymbol{\theta}}\right),
\end{eqnarray*}
where $$A\left(\overline{\boldsymbol{\theta}}\right) = \sum_{k}\ell_{k}\left(\overline{\boldsymbol{\theta}}\right), B\left(\overline{\boldsymbol{\theta}}\right) = \sum_{k}\nabla_{\boldsymbol{\theta}}\ell_{k}\left(\overline{\boldsymbol{\theta}}\right)^{\top} \text{ and } C\left(\overline{\boldsymbol{\theta}}\right) = \sum_{k}\nabla_{\boldsymbol{\theta}\boldsymbol{\theta}^{\top}}^{2}\ell_{k}\left(\overline{\boldsymbol{\theta}}\right).$$
The sums $A\left(\overline{\boldsymbol{\theta}}\right)$, $B\left(\overline{\boldsymbol{\theta}}\right)$,
and $C\left(\overline{\boldsymbol{\theta}}\right)$ are computed only once at every stage of the SMC, regardless of the number of particles. Then, for each particle, estimating $d(\boldsymbol{\theta})$ by $\widehat{d}_m(\boldsymbol{\theta}) = (n/m)\sum_j d_{u_j}(\boldsymbol{\theta})$ is computed in $O(m)$ time, and so is \eqref{eq:difference_estimator_logl} because $q(\boldsymbol{\theta})$ is $O(1)$.
We estimate $\sigma^{2}\left(\boldsymbol{\theta}\right)=\V\left(\widehat{\ell}_{m}\left(\boldsymbol{\theta}\right)\right)$ by
\[
\widehat{\sigma}_{m}^{2}\left(\boldsymbol{\theta}\right)=\frac{n^{2}}{m^2}\sum_{j=1}^{m}\left(d_{u_{j}}\left(\boldsymbol{\theta}\right)-\overline{d}_{\boldsymbol{u}}\left(\boldsymbol{\theta}\right)\right)^{2},
\]
where $\overline{d}_{\boldsymbol{u}}\left(\boldsymbol{\theta}\right)$
denotes the mean of the $d_{u_{j}}$ for the sample $\boldsymbol{u}=\left(u_{1},...,u_{m}\right)$. The estimate $\widehat{\sigma}_{m}^{2}\left(\boldsymbol{\theta}\right)$ comes at virtually no cost since it involves terms that are already computed when obtaining $\widehat{d}_{m}\left(\boldsymbol{\theta}\right)$. 


\subsection{The reweighting and resampling steps}\label{subsec:reweighting_resampling_Subsampling_SMC}
The initial particle cloud and weights are now $\left\{ \boldsymbol{\theta}_{1:M}^{\left(0\right)}, \boldsymbol{u}_{1:M}^{\left(0\right)},  W_{1:M}^{\left(0\right)}\right\}$,
obtained by generating the $\left\{ \boldsymbol{\theta}_{1:M}^{\left(0\right)}, \boldsymbol{u}_{1:M}^{\left(0\right)}\right\} $
from $p\left(\boldsymbol{\theta}\right)$ and $p\left(\boldsymbol{u}\right)$, and assigning equal weights,
i.e., $W_{1:M}^{\left(0\right)}=1/M$. The weighted particles $\left\{ \boldsymbol{\theta}_{1:M}^{\left(p-1\right)},\boldsymbol{u}_{1:M}^{\left(p-1\right)},W_{1:M}^{\left(p-1\right)}\right\} $
at the $\left(p-1\right)$st stage are a sample from $\overline{\pi}_{{p-1}} \left(\boldsymbol{\theta}, \boldsymbol{u}\right)$ and are propagated to $\overline{\pi}_{{p}}\left(\boldsymbol{\theta}, \boldsymbol{u}\right)$, by updating the weights $W_{1:M}^{\left(p\right)}=w_{1:M}^{\left(p\right)}/ \sum_{i=1}^{M}w_{i}^{\left(p\right)}$, where
$$w_i^{(p)} = W^{(p-1)}_i 
\exp\left(\left(a_p - a_{p-1}\right)\widehat{\ell}_m(\boldsymbol{\theta}^{(p-1)}_i) - \frac{1}{2}\left(a^2_p - a^2_{p-1}\right)\widehat{\sigma}_m^2(\boldsymbol{\theta}^{(p-1)}_i) \right).$$
The particles $\left\{ \boldsymbol{\theta}_{1:M}^{\left(p-1\right)},\boldsymbol{u}_{1:M}^{\left(p-1\right)}\right\}$ are then resampled using the weights $W_{1:M}^{\left(p\right)}$
to obtain the equally-weighted particles  $\left\{ \boldsymbol{\theta}_{1:M}^{\left(p\right)},\boldsymbol{u}_{1:M}^{\left(p\right)}\right\}$.

\subsection{The Markov move step\label{sub:Markov-moves}}
The Markov move step uses Hamiltonian dynamics to propose distant particle moves and data subsampling to speed up the computation of the dynamics.
Similarly to Section \ref{subsec:SMC_static_models}, the
Markov move is designed to leave each of the sequence target densities $\overline{\pi}_{{p}}\left(\boldsymbol{\theta},\boldsymbol{u}\right)$,
for $p=0,...,P,$ invariant. Algorithm~\ref{alg:AIS-BD-Markov-move} describes the Markov move step and is divided into two parts to accommodate subsampling. See \cite{Dang2017} for the details.

\begin{algorithm}[H]
\caption{Single Markov move with a kernel invariant for $\overline{\pi}_p(\boldsymbol{\theta}, \boldsymbol{u})$ in \eqref{eq:extended_target_at_sequence_p}.\label{alg:AIS-BD-Markov-move}}
\vspace*{2mm}
For $i=1,...,M$,
\begin{enumerate}
\item Sample $\boldsymbol{u}_{i}|\boldsymbol{\theta}_{i},\boldsymbol{y}$:
Propose $\boldsymbol{u_i}^{*}\sim p\left(\boldsymbol{u}\right)$,
and set $\boldsymbol{u}_{i}=\boldsymbol{u_i}^{*}$, with probability
\begin{equation}\label{eq:acc_prop_u}
\alpha_{\boldsymbol{u}}=\min\left(1, r \coloneqq \frac{\exp\left(a_{p}\widehat{\ell}_{m}\left(\boldsymbol{\theta}_{i},\boldsymbol{u_i}^{*}\right)-\frac{1}{2}a^2_{p}\widehat{\sigma}_{m}^{2}\left(\boldsymbol{\theta}_{i},\boldsymbol{u_i}^{*}\right)\right)}{\exp\left(a_{p}\widehat{\ell}_{m}\left(\boldsymbol{\theta}_{i},\boldsymbol{u}_{i}\right)-\frac{1}{2}a^2_{p}\widehat{\sigma}_{m}^{2}\left(\boldsymbol{\theta}_{i},\boldsymbol{u}_{i}\right)\right)}\right),
\end{equation}

The proposal $\boldsymbol{u_i}^{*}$ is independent
of the current value of $\boldsymbol{u}_{i}$, so the difference between
the log of the numerator and log of the denominator of the ratio $r$ in \eqref{eq:acc_prop_u} can be highly variable. This move might get stuck when the denominator is significantly
overestimated. A remedy is to induce a high correlation between the log of the estimated annealed likelihood at the current and proposed draws in \eqref{eq:acc_prop_u}. This can be achieved either through correlating the $\boldsymbol{u}$
as in \citet{Deligiannidis2017} (see \citealt{Quiroz2018} for discrete $\boldsymbol{u}$) or by block updates of $\boldsymbol{u}$ as in \citet{Tran:2016, quiroz2018block}. We implement the
block updates with $G$ blocks, which gives an approximate correlation $1-\frac{1}{G}$.

\item Sample $\boldsymbol{\theta}_{i}|\boldsymbol{u}_{i},\boldsymbol{y}$:
Given a subset of data $\boldsymbol{u}_{i}$, we move the particle
$\boldsymbol{\theta}_{i}$ using a Hamiltonian Monte Carlo (HMC) proposal in a Metropolis-Hastings (MH) algorithm.
This becomes a standard HMC move for a given subset $\boldsymbol{u}$.
\end{enumerate}
Note that the above is a Gibbs update of $\boldsymbol{\theta}_i, \boldsymbol{u}_i | \boldsymbol{y} $. The MH within Gibbs performed in Step 1. is valid \citep{johnson2013component} and so is the HMC within Gibbs \citep{Neal:2011} in Step 2. Therefore, this kernel has $\overline{\pi}_p(\boldsymbol{\theta}, \boldsymbol{u})$ as its invariant distribution. \citet{Dang2017} previously proposed an MCMC version of this algorithm.
\end{algorithm}

Algorithm \ref{alg:AIS-Algorithm-bigdata} summarizes our approach. We follow \cite{buchholz2018adaptive} who develop a tuning procedure for the mass matrix, the step size and the number of leapfrog steps within an SMC framework. The number of Markov moves $R$ is tuned by increasing it until 90$\%$ of the product of componentwise autocorrelation of the particles drops below a threshold; see \cite{buchholz2018adaptive} for more details.

\begin{algorithm}[H]
\caption{Subsampling Sequential Monte Carlo \label{alg:AIS-Algorithm-bigdata}}

\begin{enumerate}
\item Sample the particles $\left\{ \boldsymbol{\theta}_{i}^{\left(0\right)},\boldsymbol{u}_{i}^{\left(0\right)}\right\} $ from the prior densities $p\left(\boldsymbol{\theta}\right)$ and $p\left(\boldsymbol{u}\right)$ and give all particles equal
weights, $W_{i}=1/M$, $i=1,...,M$.
\item While the tempering sequence $a_{p}\neq1$ do
\begin{enumerate}
\item Set $p\leftarrow p+1$
\item Find $a_{p}$ adaptively to maintain the ESS around $\mathrm{ESS}_{\mathrm{target}}$ (Section \ref{subsec:SMC_efficiency}).
\item Reweighting: compute the unnormalized weights
\begin{eqnarray*}
w_{i}^{\left(p\right)} &= &W_{i}^{\left(p-1\right)}\frac{\eta_{a_{p}}\left(\boldsymbol{\theta}_{i}^{(p-1)},\boldsymbol{u}_{i}^{(p-1)}\right)}{\eta_{a_{p-1}}\left(\boldsymbol{\theta}_{i}^{(p-1)},\boldsymbol{u}_{i}^{(p-1)}\right)}\\
& = & W_{i}^{\left(p-1\right)}\exp\left(\left(a_p - a_{p-1}\right)\widehat{\ell}_m(\boldsymbol{\theta}^{(p-1)}_i) - \frac{1}{2}\left(a^2_p - a^2_{p-1}\right)\widehat{\sigma}_m^2(\boldsymbol{\theta}^{(p-1)}_i) \right),
\end{eqnarray*}
and normalize as $W_{i}^{\left(p\right)}=w_{i}/\sum_{i^\prime=1}^{M}w_{i^\prime}$,
$i=1,...,M$.
\item Compute $\overline{\boldsymbol{\theta}}$ as $\overline{\boldsymbol{\theta}}=\sum_{i=1}^{M}W_{i}^{\left(p\right)}\boldsymbol{\theta}_{i}^{\left(p - 1\right)}$
\and and then obtain $$\sum_{k=1}^{n}\ell_{k}\left(\overline{\boldsymbol{\theta}}\right),\,\,
\sum_{k=1}^{n}\nabla_{\boldsymbol{\theta}}\ell_{k}\left(\overline{\boldsymbol{\theta}}\right),\,\, \sum_{k=1}^n\nabla_{\boldsymbol{\theta}\boldsymbol{\theta}^{\top}}^{2}\ell_{k}\left(\overline{\boldsymbol{\theta}}\right)$$
and the mass matrix $\boldsymbol{H}=\boldsymbol{\Sigma}^{-1}\left(\overline{\boldsymbol{\theta}}\right)$, where $\Sigma$ is the sample covariance matrix of current particles.
\item Resample the particles $\left\{ \boldsymbol{\theta}_{i}^{\left(p-1\right)},\boldsymbol{u}_{i}^{\left(p-1\right)}\right\} _{i=1}^{M}$
using the weights $\left\{ W^{(p)}_{i}\right\} _{i=1}^{M}$ to obtain resampled particles $\left\{ \boldsymbol{\theta}_{i}^{\left(p\right)},\boldsymbol{u}_{i}^{\left(p\right)}\right\} _{i=1}^{M}$  and set $W^{(p)}_i = 1/M$.
\item Apply $R$ Markov moves to each particle $\boldsymbol{\theta}_{i}^{\left(p\right)},\boldsymbol{u}_{i}^{\left(p\right)}$ using Algorithm \ref{alg:AIS-BD-Markov-move}.
\end{enumerate}
\end{enumerate}
\end{algorithm}

\subsection{Marginal likelihood estimation \label{sub:Estimating-Marginal-Likelihood}}
Our approach naturally extends that of Section \ref{subsec:marginal_likelihood_SMC} by considering the augmented target density $\overline{\pi}_p(\boldsymbol{\theta}, \boldsymbol{u}) $ in \eqref{eq:extended_target_at_sequence_p}. Define $$\gamma_p(\boldsymbol{\theta}, \boldsymbol{u})=\frac{\eta_p(\boldsymbol{\theta}, \boldsymbol{u})}{\eta_{p-1}(\boldsymbol{\theta}, \boldsymbol{u})}.$$
Then
\begin{eqnarray*}
\int_{\mathcal{U}} \int_{\boldsymbol{\Theta}} \gamma_p(\boldsymbol{\theta}, \boldsymbol{u})\pi_{p-1}(\boldsymbol{\theta}, \boldsymbol{u})d\boldsymbol{\theta}d\boldsymbol{u} & = & \int_{\mathcal{U}} \int_{\boldsymbol{\Theta}} \frac{\eta_p(\boldsymbol{\theta}, \boldsymbol{u})}{\eta_{p-1}(\boldsymbol{\theta}, \boldsymbol{u})} \frac{\eta_{p-1}(\boldsymbol{\theta}, \boldsymbol{u})}{Z_{p-1}}p(\boldsymbol{\theta})p(\boldsymbol{u})d\boldsymbol{\theta}d\boldsymbol{u} \\
~ & = & \frac{Z_p}{Z_{p-1}}.
\end{eqnarray*}
Thus, if $\left\{ \boldsymbol{\theta}_{1:M}^{\left(p-1\right)},\boldsymbol{u}_{1:M}^{\left(p-1\right)},W_{1:M}^{\left(p-1\right)}\right\} $ at the $(p-1)$st sequence is an approximate sample from $\overline{\pi}_{a_{p-1}}\left(\boldsymbol{\theta},\boldsymbol{u}\right)$,
we estimate the ratio $Z_p/Z_{p-1}$ by
\[
\widehat{\frac{Z_{{p}}}{Z_{{p-1}}}}=\sum_{i=1}^{M}W_{i}^{\left(p-1\right)}\frac{\eta_{{p}}\left(\boldsymbol{\theta}_{i}^{\left(p-1\right)},\boldsymbol{u}_{i}^{\left(p-1\right)}\right)}{\eta_{{p-1}}\left(\boldsymbol{\theta}_{i}^{\left(p-1\right)},\boldsymbol{u}_{i}^{\left(p-1\right)}\right)},
\]
and the marginal likelihood estimate is obtained using this expression in \eqref{eq:marginal_likelihood_estimate}.

\subsection{Efficient memory management by data subsampling} \label{subsec:MemoryEfficiencySubsamplingSMC}
We now explain in detail how data subsampling helps to parallelize the computing in terms of efficient memory utilization. Suppose first that we perform standard SMC (using all the data) and that we parallelise using $N$ workers, so that each worker deals, on average, with $M/N$ particles. Then, for each stage $p$, the computations performed for each particle require repeated likelihood evaluations (using all $n$ data) when applying $R$ Markov move steps. Hence, each worker needs to have access to the full dataset.

Suppose now that we use our data subsampling approach in the same setting using $M/N$ particles for each of the $N$ workers. Then, at the beginning of each stage $p$ of the algorithm, we still require a full data evaluation for computing $A(\overline{\bs \theta}), B(\overline{\bs \theta})$ and $C(\overline{\bs \theta})$ in Section \ref{subsec:EfficientEstimator}. However, at each $p$, we can now subsample the data according to $\bs{u}_i^{(p)}$ for each particle and subsequently perform the $R$ Markov move steps, which now require repeated evaluations of the estimated annealed likelihood (using $m \ll n$ observations) and in addition $A(\overline{\bs \theta}), B(\overline{\bs \theta})$ and $C(\overline{\bs \theta})$. Now each worker needs to have access only to the subsampled dataset, as well as $A(\overline{\bs \theta}), B(\overline{\bs \theta})$ and $C(\overline{\bs \theta})$. However, these are only summaries of the full dataset and are therefore very memory efficient. 

We are aware that parallelization of SMC methods is not straightforward to do efficiently when resampling occurs often \citep{murray2016parallel,lee2010utility}. We note that in all our applications the number of annealing steps is relative small and therefore resampling does not really affect the efficiency of our algorithm. In applications where resampling occurs more frequently, both SMC methods can benefit from the ideas in \cite{heine2019parallelizing} and \cite{guldas2015practical}. Moreover, the reweighting and the computationally expensive Markov move steps of our algorithm are easily parallelised for each SMC sample because the computations required for each sample are independent of those of the other samples. Subsampling therefore does not affect the parallelisation of the algorithm because only the part of the data specified by the particles $u_i$ are sent to each worker and the $u_i$ are independent of each other.



\section{Evaluations \label{sec:Evaluations}}
\subsection{Experiments\label{subsec:Experiments}}
We now evaluate the methodology through the following experiments.
\begin{itemize}
\item \textit{Experiment 1: Evaluating the usefulness of the Hamiltonian Monte Carlo kernel.}\\
We show the effectiveness of a HMC kernel for the Markov move step compared to random walk and MALA kernels.

\item \textit{Experiment 2: Evaluating the speed and the accuracy of the marginal likelihood and the approximate posterior density when the posterior is unimodal.}\\
We show that the subsampling approach is accurate by comparing the estimates of the marginal likelihood and posterior densities to those obtained by the full data SMC (representing the gold standard). 

\item \textit{Experiment 3: Evaluating the speed and the accuracy of the marginal likelihood and the approximate posterior density when the posterior is non-Gaussian.}\\
We use the subsampling approach when the posterior is bimodal or skewed and show that the method still performs well.

\item \textit{Experiment 4: Evaluating the effect of the accuracy of the control variate.} \\
We show that the subsampling approach can be made faster by using a first order control variate instead of the second order alternative. This experiment also shows the effect of inaccurate likelihood estimates on the performance of our method. 

\end{itemize}

All the SMC algorithms are tuned as in \cite{buchholz2018adaptive} using $280$ particles, a choice motivated by our cluster with $28$ cores with each core dealing (on average) with $10$ particles.  The only exception is the first scenario in Experiment 3 where we use $420$ particles for both algorithms to better capture the multimodal posterior. We repeated each experiment $10$ times to compute the standard error of the log marginal likelihood estimator. Experiments 1, 2 and the bankruptcy application in Section \ref{sec:RealApplication} were done using the Australia NCI High Performance Computing System Raijin\footnote{https://nci.org.au/our-systems/hpc-systems}. Experiments 3 and 4 were done using the University of New South Wales computational cluster Katana\footnote{https://research.unsw.edu.au/katana}.




	

We remark that the choice of priors can affect the computational efficiency of SMC methods. In general, a prior that resembles the likelihood requires less tempering steps. However, this is unlikely to influence the comparison between Subsampling SMC and SMC, which is our primary concern. 
 
\subsection{Experiment 1: Evaluating the Markov move kernel\label{subsec:Experiment1}}
We first consider a logistic regression to evaluate how effectively the Hamiltonian Monte Carlo Markov move step is compared to the random walk and MALA kernels. The model for the response $y_{i}\in\left\{ 0,1\right\}$ given a $d\times 1$ set of covariates and parameters is
\begin{equation*}
p\left(y_{i}|\boldsymbol{x}_{i},\boldsymbol{\theta}\right)=\frac{\exp\left(y_i x_i^{\top}\boldsymbol{\theta}\right) }{1+\exp\left(\boldsymbol{x}_{i}^{\top}\boldsymbol{\theta}\right)}.
\end{equation*}
We fit this model to the HIGGS dataset \citep{Baldi2014}, having $n=11{,}000{,}000$ observations and $28$ covariates. The response is ``detected particle'' and $21$ of the covariates are kinematic properties measured by particle detectors, while $7$ are high-level features to capture non-linearities. This means that $d =29$, including the intercept. We take the prior $\boldsymbol{\theta} \sim \mathcal{N}(\mathbf{0},\mathbf{I}_{d})$, where $\mathbf{I}_{d}$ is the $d \times d$ identity matrix and follow \cite{buchholz2018adaptive} to set the tuning parameters, including the number of Markov moves $R$. The mass matrix in both HMC and MALA is $\widehat{\Sigma}^{-1}$,  which is the estimated inverse covariance matrix of the tempered posterior. We note that each step in the sequence has a corresponding estimate of this inverse covariance matrix, obtained using the corresponding particles from that step. For the random walk, the optimal scaling $(2.38^2/d) \widehat{\Sigma}$ \citep{Roberts:1997} resulted in numerical errors, so that we decreased it by a factor of $10$. 

Table \ref{tab:Table_HIGGS_different_kernels} summarizes the results obtained using the second order control variate in Section \ref{subsec:EfficientEstimator}. The log-likelihood estimator has $m=5{,}000$ subsamples and the block-pseudo marginal is carried out using $G=100$.  Clearly, the Hamiltonian approach is computationally faster because it needs to take a smaller number of Markov steps $R$. The table also shows that the log of the estimate of marginal likelihood is very similar for all methods. The rest of the article uses the HMC kernel.
\begin{table}
 	\centering
 	  	\caption{Comparing the performances of three kernels for the Markov move, Hamiltonian Monte Carlo (HMC), Metropolis Adjusted Langevin Algorithm (MALA) and Random Walk (RW). The table shows the log of the estimate of the marginal likelihood (with standard error in brackets), the CPU time, the number of annealing steps $P$ (tuned to maintain $ESS \approx 0.8M$) and the number of Markov moves $R$ (tuned as in \citealp{buchholz2018adaptive}). The results are for the logistic regression model estimated using the HIGGS data and $M=280$ particles. All methods use the second order control variate in Section \ref{subsec:EfficientEstimator}. The results are averaged over $10$ runs, which are used to compute the standard error of the estimator.}

 	\fontsize{11}{11}\selectfont
 	\begin{tabular}{*5c} \toprule
 		& & &\\
 		 & log marginal likelihood  & CPU time (hrs)& $P$ & $R$ \\ [5pt]\midrule
 		& & \\
 		HMC&  -7,013,460.90 &2.31  & 106 & 5 \\ 
 		& \tiny{(0.32)} & & & \\ [5pt]
 		MALA & -7,013,462.49& 4.77  & 106&  20 \\
 		& \tiny{(0.26)} & & & \\ [5pt]
 		RW &  -7,013,461.43  &33.43  & 106 & 200\\
 		& \tiny{(0.32)} & & & \\[5pt]
 		\bottomrule
 	\end{tabular}	
 	\vspace{3mm}
 	\label{tab:Table_HIGGS_different_kernels}
 \end{table}

\subsection{Experiment 2: Evaluating speed and accuracy of Subsampling SMC on unimodal targets\label{subsec:Experiment2}}

This section compares Subsampling SMC with full data SMC. Such a comparison is infeasible for the full HIGGS dataset because it is too large; 
the full dataset needs to be available at each worker (we use $28$) as explained in Section \ref{subsec:MemoryEfficiencySubsamplingSMC},  in order to compute the likelihood together with its gradient and Hessian, which would quickly consume
the RAM of the computer. Instead, we 
consider the following two models. 

\textbf{Student-t regression.} 
We consider a univariate Student-t regression
\begin{equation*}
y_i = \boldsymbol{x}_{i}^{\top}\boldsymbol{\theta} + e_i, \, e_i \sim t_5,
\end{equation*}
where $t_5$ is the Student-t distribution with $5$ degrees of freedom. We generated a simulated dataset with $n=500{,}000$ observations 
and $d = 50$ covariates. The covariates were generated so that their marginal variances are $1$ and their pairwise correlations are $0.9$. 
The parameters $\boldsymbol{\theta}$ were simulated independently from a $\mathrm{Uniform}(-5,5)$ distribution;  
the prior for $\boldsymbol{\theta}$ is  $\mathcal{N}(\mathbf{0}, 10 \mathbf{I}_{d})$.

\textbf{Poisson regression.} 
We also considered a Poisson regression, where the univariate $y$ follows a Poisson distribution with an expectation that is log-linear, i.e. $$y_i | \boldsymbol{x}_i \sim \mathrm{Poisson}(\exp(\boldsymbol{x}_{i}^{\top}\boldsymbol{\theta})).$$ We generated $n=200{,}000$ observations with $d = 30$ covariates, $29$ of them simulated from $\boldsymbol{x}_i \sim \mathcal{N}(\mathbf{0},  \mathbf{I}_{29})$ and the last one is $1$. The parameters are simulated independently from $\mathrm{Uniform}(-0.2,0.2)$ and are assigned the prior $\boldsymbol{\theta} \sim \mathcal{N}(\mathbf{0}, 0.1 \mathbf{I}_{d})$. We found that both Subsampling SMC and SMC were particularly sensitive to the prior choice for the Poisson regression, resulting in numerical overflow for priors that were too diffuse.

\begin{figure}
	\centering
	\includegraphics[width=0.9\linewidth]{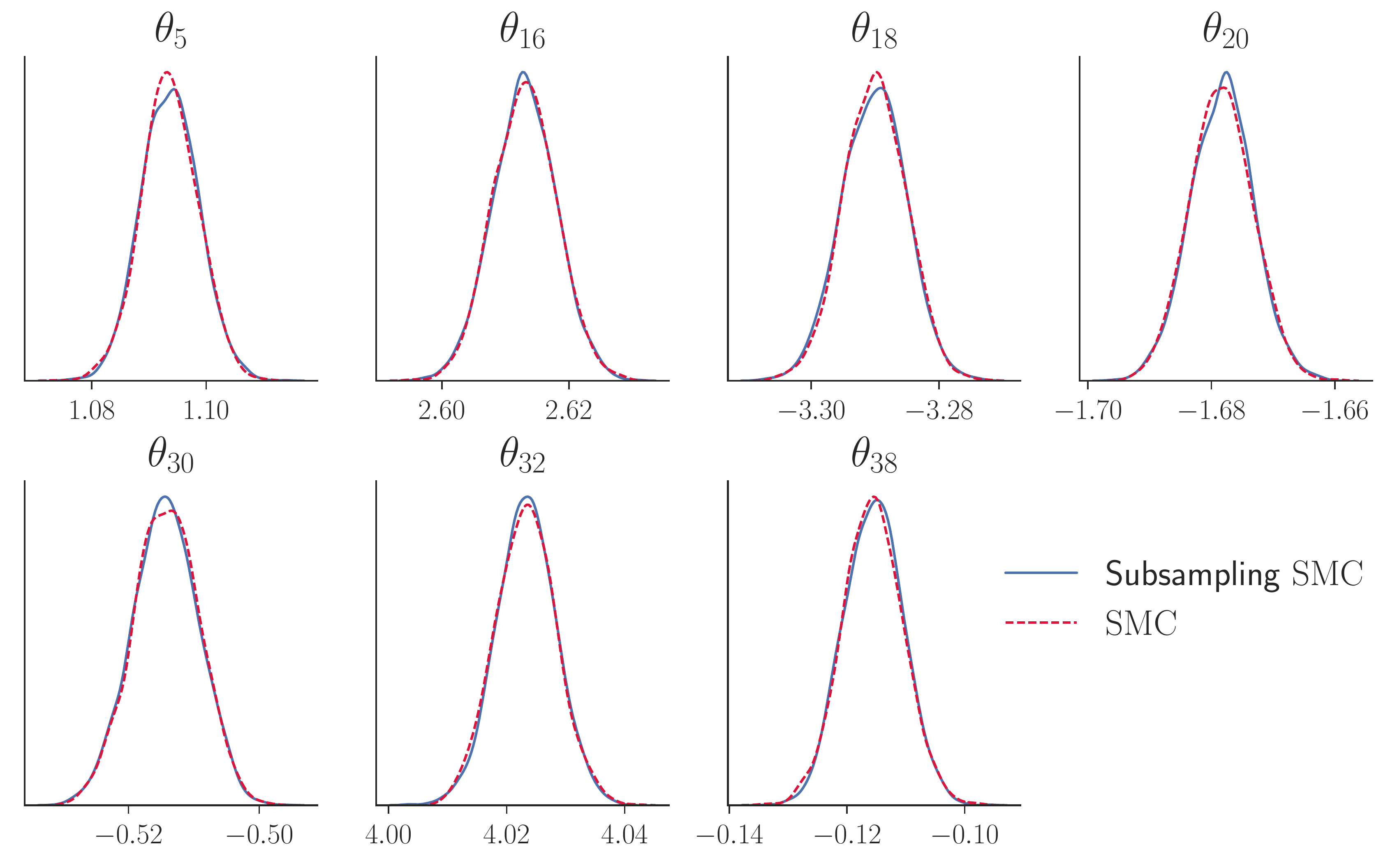}
	\caption{Kernel density estimates of a subset of the marginal posterior densities of $\boldsymbol{\theta}$ for the Student-t regression model with simulated data. The density estimates are obtained by full data SMC and Subsampling SMC.}
	\label{fig:comparekdetreg}
\end{figure}

\begin{figure}
	\centering
	\includegraphics[width=0.8\linewidth]{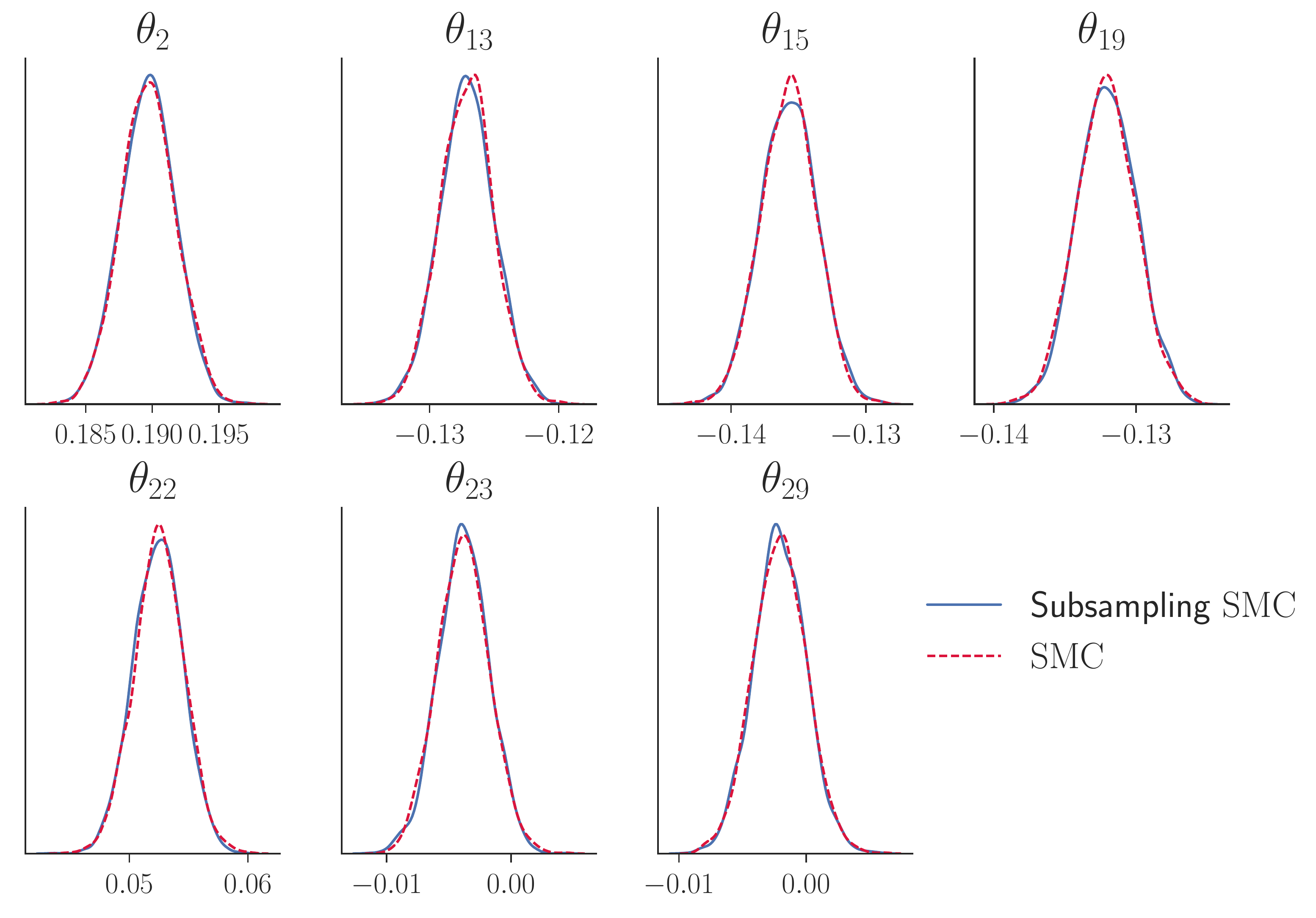}
	\caption{Kernel density estimates of a subset of the marginal posterior densities of $\boldsymbol{\theta}$ for the Poisson regression model with simulated data. The density estimates are obtained by full data SMC and Subsampling SMC.}
	\label{fig:kdepoisson}
\end{figure}

For both examples, we used $G=100$ blocks and the second order Taylor series control variates and set $m$ to correspond to a sample fraction of about $0.0025$. Table \ref{tab:Poisson_and_Student_t} summarizes the results and shows that the subsampling approach is about 6.5 to 10.5 times faster and, moreover, confirms the accuracy of the marginal likelihood estimate of our method. 
\begin{table}
	\centering
	 	  	\caption{Comparing the performances of Subsampling SMC and full data SMC. The table shows the log of the estimate of the marginal likelihood (with standard error in brackets), the CPU time, the number of annealing steps $P$ (tuned to maintain $ESS \approx 0.8M$) and the number of Markov moves $R$ (tuned as in \citealp{buchholz2018adaptive}). The results are for the Student-t regression and Poisson regression models estimated using the simulated datasets. We use $M=280$ particles. All methods use the second order control variate in Section \ref{subsec:EfficientEstimator}. The results are averaged over $10$ runs, which are used to compute the standard error of the estimator.}
	\fontsize{11}{11}\selectfont
	\begin{tabular}{*5c} \toprule
		& & &\\
		& log marginal likelihood  & CPU time (hrs)& $P$  & $R$ \\ [5pt]\midrule
		\underline{\textbf{Student-t regression}} & & \\
		 & & \\  
		 ($n = 500{,}000, m = 1{,}200$) & & \\
		 & & \\  
		Full data SMC & -815,775.82  &5.92  & 126 & 4 \\
		& \tiny{(0.39)}  & &  \\[5pt]
		Subsampling SMC &  -815,773.49 & 0.57  & 127 & 4 \\ 
		& \tiny{(0.59)}  & & & \\ [5pt]
		& & \\
		\underline{\textbf{Poisson regression}} & & \\
		 & & \\  
		 ($n = 200{,}000, m = 500$) & & \\
		 & & \\  

		SMC &-260,888.69 &0.94  & 80 & 4 \\
		& \tiny{(1.40)}  & &  \\[5pt]
		Subsampling SMC &  -260,887.87 &0.14  & 80 & 5 \\ 
		& \tiny{(0.27)} & & & \\ [5pt]
		\bottomrule
	\end{tabular}	
	\vspace{3mm}

	\label{tab:Poisson_and_Student_t}
\end{table}
Finally, Figures \ref{fig:comparekdetreg} and \ref{fig:kdepoisson} show that the marginal posterior densities are very well approximated for both the Student-t regression and the Poisson regression; the same accuracy was obtained for all parameters (not shown).

\subsection{Experiment 3: Evaluating speed and accuracy of Subsampling SMC on non-Gaussian targets \label{subsec:Experiment3}}
To evaluate the performance of Subsampling SMC when the posterior is non-Gaussian, we consider the fixed effects model 
\begin{equation*}
	y_{ij}= \alpha_i +  \boldsymbol{x}_{ij}^{\top}\boldsymbol{\beta} + e_{ij},\quad i = 1,\dots, n, \quad j = 1,\dots, n_i, \quad e_{ij} \sim \mathcal{N}(0,\sigma^2_i).
	\end{equation*}
For simplicity, we set $\sigma^2_i= 1$ for all 10 individuals; two different scenarios  are used for the individual fixed effects $\alpha_i$; 
a)  a mixture of normals prior for the $\alpha_i$; b) a truncated normal prior for the $\alpha_i$. 
	For each scenario, we generated a dataset of $n=10$ individuals, with $n_1=\dots =n_5 = 20$ observations and $n_6 = \cdots = n_{10} = 50{,}000$ 
	observations. 	The covariates were generated independently from $\mathcal{N}(0,1)$; the $\boldsymbol{\beta}$ parameters were
	generated from $\mathcal{N}(0,2^2\mathbf{I}_{10})$.  
The prior for $\boldsymbol{\beta}$ in both scenarios is  $\mathcal{N}(0,3^2\mathbf{I}_{10})$.
\paragraph{Mixture of normals prior}

The first prior is motivated by a variable selection scenario, where some coefficients may be 0 or very close to 0 and we would like 
the posterior to set these close to zero.  
In this experiment, the first 5 individual fixed effects $\alpha_i$ were generated from $\mathcal{N}(0.5,0.05^2)$ and the rest from $\mathcal{N}(0.5,0.2^2)$. For each of the fixed effects $\alpha_i$ we used a mixture of normals prior
	$$p(\alpha_i|w,\sigma_1,\sigma_2) = w\phi(\alpha_i|\sigma_1^2)+(1-w)\phi(\alpha_i|\sigma_2^2), \quad i = 1,\dots,n;$$
$\phi(\cdot|\sigma^2)$ is the density of the normal distribution with mean 0 and variance $\sigma^2$, and we set $w=0.8$, $\sigma_1 = 0.1$ and $\sigma_2=3.5$.

Even though the likelihood for each individual is likely to be unimodal, the prior leads to more complicated posteriors for those 
individual effects that correspond to subjects with a small number of observations. The likelihood is
\begin{equation}
    L(\boldsymbol{\alpha},\boldsymbol{\beta}) 
    = \prod_{i=1}^n p(\boldsymbol{y}_i|\alpha_i,\boldsymbol{\beta}), \quad \text{where} \quad 
   p(\boldsymbol{y}_i|\alpha_i,\boldsymbol{\beta}) = \prod_{j = 1}^{n_i}p(y_{ij}|\alpha_i,\boldsymbol{\beta})
\end{equation}
is the likelihood for subject $i$. 
The likelihood  $L(\boldsymbol{\alpha},\boldsymbol{\beta})$ and the annealed likelihood are estimated by
estimating the individual likelihoods with subsampling. The subsample size is $m = 5$, with no blocking for the first 5 individuals and $m = 100$ with $G=100$ blocks for the remaining 5 individuals. In practice, it is unnecessary to estimate the likelihood for the individuals with few observations since it is  relatively cheap computationally to evaluate their full likelihood; however,  we do so  in our experiment
to gain more knowledge about the effect of subsampling. 

\begin{figure}
	\centering
	\includegraphics[width=0.9\linewidth]{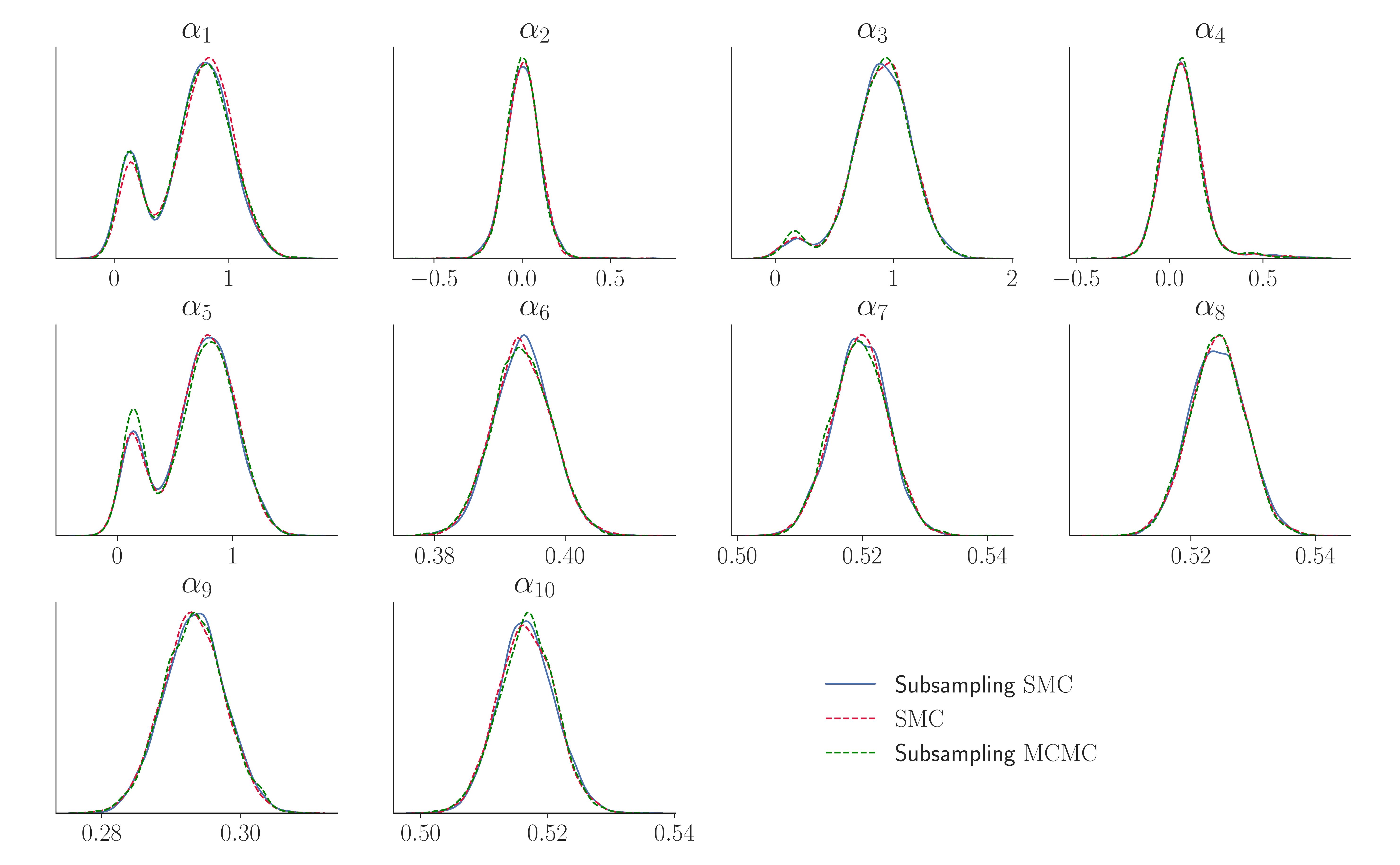}
	\caption{Kernel density estimates of a subset of the marginal posterior densities of $\boldsymbol{\alpha}$ for the fixed effects model with mixture of normals priors, using simulated data. The density estimates are obtained by full data SMC, Subsampling SMC and Subsampling MCMC.}
	\label{fig:comparekdemultimode}
\end{figure}

We ran Subsampling SMC with second order Taylor series expansions in both scenarios. Table \ref{tab:FE_example} summarizes the results and shows that Subsampling SMC produces similar results to full data SMC but is about 9 times faster. All SMC methods require the maximum number of Markov moves at most temperatures, indicating that the posterior is challenging to explore. Figure \ref{fig:comparekdemultimode} shows that even when some of the marginal posteriors ($\alpha_1, \alpha_3$ and $\alpha_5$) are bimodal, Subsampling SMC is able to capture that and gives the same approximation as full data SMC. As a comparison, we also include in the figure the result from running $10{,}000$ post burn-in iterations of Subsampling MCMC. It is well known that conventional MCMC methods may not be able to sample efficiently from multimodal targets, and in this experiment Subsampling MCMC can detect the posterior modes, but there is still some visible discrepancy between its result and that of full data SMC. We do not show the marginal posterior densities of $\boldsymbol{\beta}$ which appear to be Gaussian, but confirm that both methods give similar results.

Our method works in this example because the bimodality is caused by the prior and not the likelihood;  
if the bimodality was caused by the likelihood, different  control variates would be necessary since our control variates assume
the log-density is quadratic $\v \theta$. We leave the development of more flexible control variates for Subsampling SMC for future research.

\paragraph{Truncated normal prior}

The second scenario is motivated by situations in which there is strong prior knowledge 
that the coefficients are positive. To create such a situation,
the fixed effects $\alpha_i$ were generated from a truncated normal distribution $\mathcal{TN}(0.1,0.1^2)$.  
We assigned the prior $\alpha_i \sim \mathcal{TN}(0,3^2), i = 1,\dots,10$ to the individual fixed effects 
to reflect this prior knowledge.
The subsample size is $m = 20$ (all observations) with no blocking for the first 5 individuals and $m = 200$ with $G=100$ blocks for the $6^{th}$ individual. The remaining 4 individuals have $m= 100$ with $G= 100$. Note that the subsample size affects the variance of the log-likelihood estimator and hence the accuracy of our method; see \cite{Quiroz2018} and \cite{Dang2017} for further discussion. 
Section \ref{subsec:Experiment4} discusses the results when  a smaller subsample size is used for this model. 

\begin{figure}
	\centering
	\includegraphics[width=0.9\linewidth]{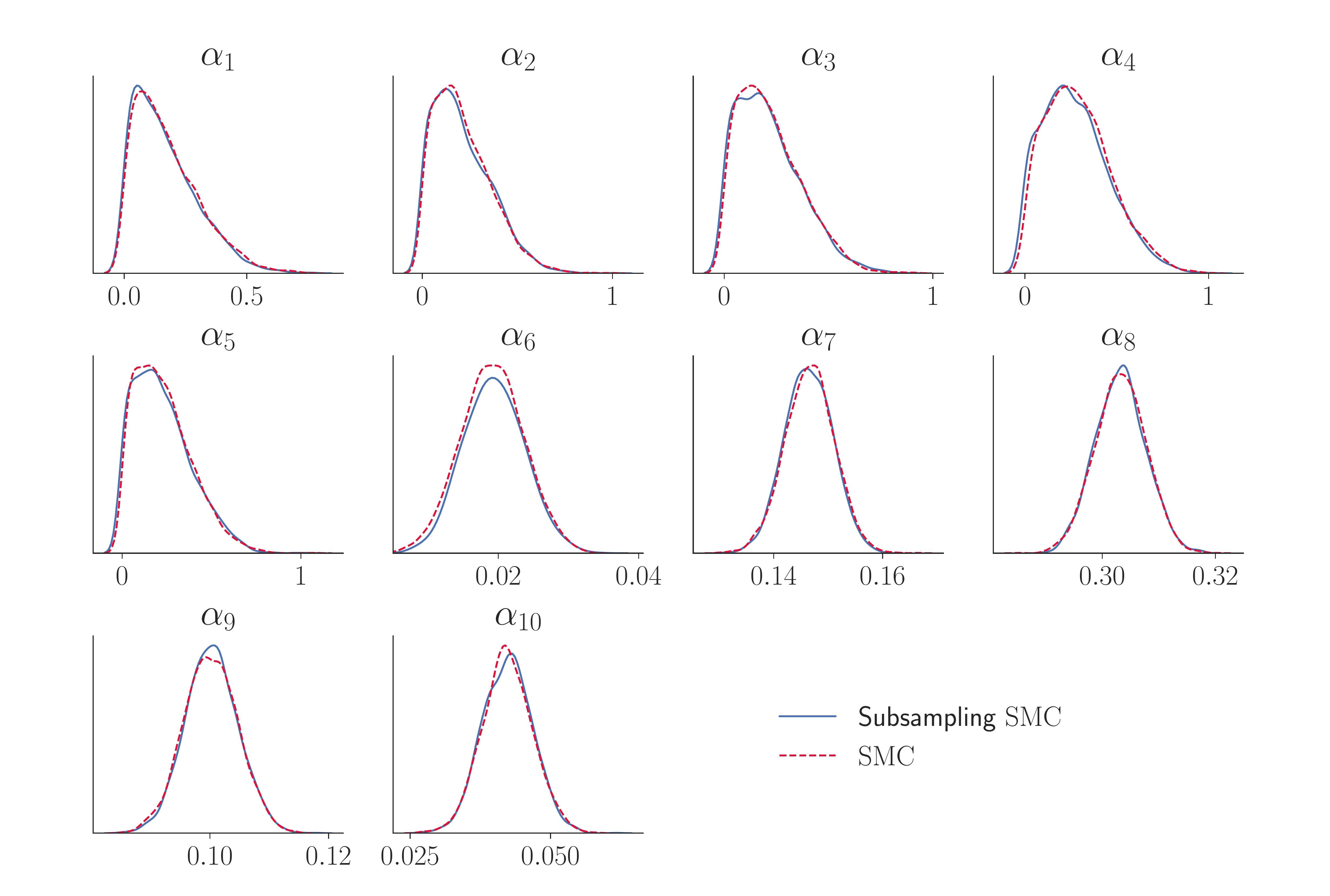}
	\caption{Kernel density estimates of a subset of the marginal posterior densities of $\boldsymbol{\alpha}$ for the fixed effects model with truncated normal priors, using simulated data. The density estimates are obtained by full data SMC and Subsampling SMC.}
	\label{fig:FEm20}
\end{figure}

Table \ref{tab:FE_example} and Figure \ref{fig:FEm20} summarize the results of full data SMC and subsampling SMC. For this example, the truncated normal prior makes the posterior of $\alpha_1,\dots ,\alpha_5$ skewed. These are the effects corresponding to the subjects with few observations. Subsampling SMC seems to experience some difficulties with this challenging target, which is shown by the slightly higher $P$ and $R$ values compared to full data SMC. However our method is still slightly faster and produces posterior estimates similar to the full data SMC (see Figure \ref{fig:FEm20}). We do not show the marginal posterior densities of $\boldsymbol{\beta}$ which appear to be Gaussian, but  both methods gave similar results.

\begin{table}
\centering
	 	  	\caption{Comparing the performances of Subsampling SMC and full data SMC. The table shows the log of the estimate of the marginal likelihood (with standard error in brackets), the CPU time, the number of annealing steps $P$ (tuned to maintain $ESS \approx 0.8M$) and the number of Markov moves $R$ (tuned as in \citealp{buchholz2018adaptive}, and the maximum number of Markov moves at each temperature is set to be 100). The results are for the fixed effects model estimated using the simulated datasets. All methods use the second order control variate in Section \ref{subsec:EfficientEstimator}. The results are averaged over $10$ runs, which are used to compute the standard error of the estimator.}
	\fontsize{11}{11}\selectfont
\begin{tabular}{*5c} \toprule
		& & &\\
		& log marginal likelihood  & CPU time (hrs)& $P$  & $R$ \\ [5pt]\midrule
		
		\underline{\textbf{Mixture of normals priors}} 
		 & & \\  [5pt]
		 ($M= 420$) & & \\
		 & & \\
		Full data SMC & -354,914.89  &14.36  & 81 & 99 \\
		& \tiny{(0.78)}  & &  \\[5pt]
		Subsampling SMC &  -354,915.22 & 1.60  & 81 & 100 \\ 
		& \tiny{(1.18)}  & & & \\ [5pt]
		& & \\
		\underline{\textbf{Truncated normal priors}} 
		 & & \\  [5pt]
		 ($M= 280$) & & \\
		 & & \\
		Full data SMC &-354,445.12   & 0.74 & 79 & 5 \\
		& \tiny{(0.26)}  & &  \\[5pt]
		Subsampling SMC &-354,444.04   & 0.68  & 88 & 20 \\ 
		& \tiny{(2.2)}  & & & \\ [5pt]
\bottomrule
	\end{tabular}	
	\vspace{3mm}

	\label{tab:FE_example}
\end{table}

\subsection{Experiment 4: Evaluating the effect of the control variate\label{subsec:Experiment4}}
\begin{figure}
	\centering
	\includegraphics[width=0.9\linewidth]{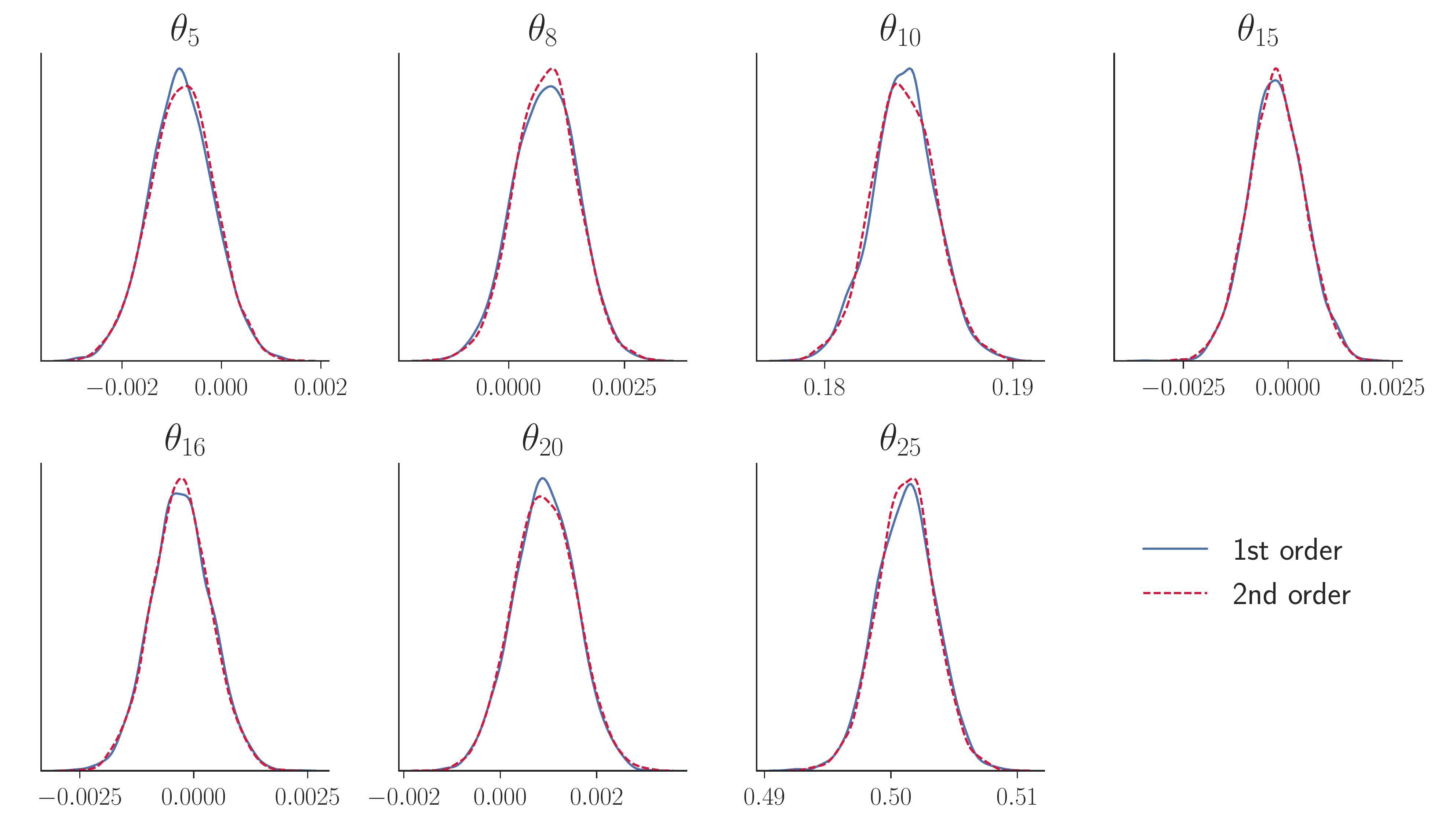}
	\caption{Kernel density estimates of a subset of the marginal posterior densities of $\boldsymbol{\theta}$ for the logistic model with the HIGGS data. The density estimates are both obtained by Subsampling SMC, using different control variates based on a 1st and 2nd order Taylor series expansion as explained in Section \ref{subsec:EfficientEstimator}.}
	\label{fig:comparecv}
\end{figure}

The results above show that the subsampling  approach accurately estimates the marginal likelihood and marginal posterior densities using a second order Taylor series expansion. We now study robustness of the results to the quality of the control variates. The first study
uses first order Taylor expansions for the control variates for subsampling applied to the 
logistic regression for the HIGGS dataset in Section \ref{subsec:Experiment1}. Table \ref{tab:Experiment4} summarizes the results, and confirms that the marginal likelihood estimator remains accurate, and is five times faster than using the second order control variates. 
Figure \ref{fig:comparecv} shows that the marginal posterior densities remain accurate, we have confirmed similar accuracy for all the parameters. 


We now present an example where inaccurate likelihood estimates lead to a biased result. We consider again the fixed effects model described in Section \ref{subsec:Experiment3} with the individual effects $\alpha_i$ having a truncated normal prior, $p(\alpha_i) \sim \mathcal{TN}(0,3^2)$, $m =5$ is used for the first 5 individuals and $m= 100$ with $G=100$ for the remaining 5 individuals. 

\begin{figure}
	\centering
	\includegraphics[width=0.9\linewidth]{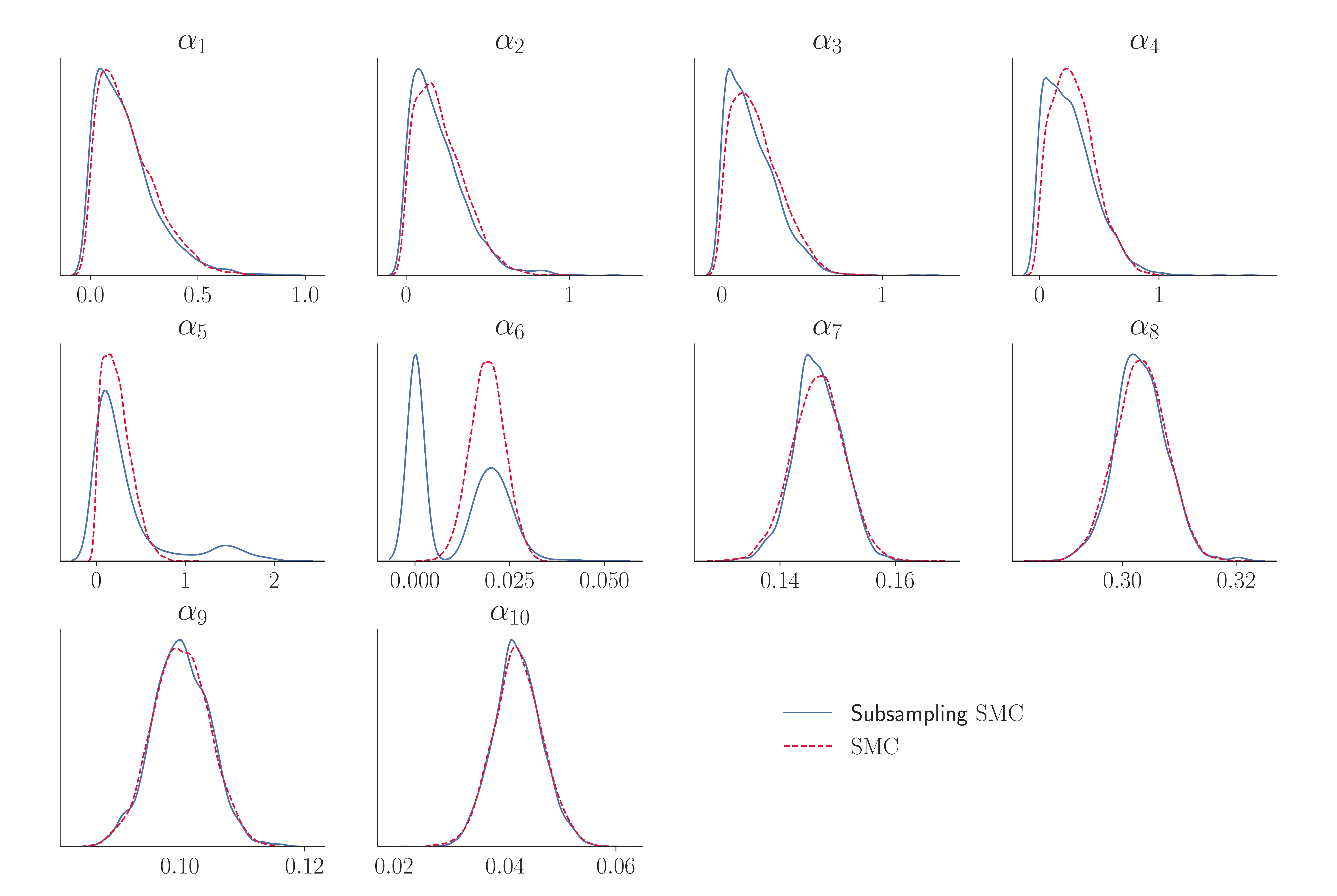}
	\caption{Kernel density estimates of a subset of the marginal posterior densities of $\boldsymbol{\alpha}$ for the fixed effects model with truncated normal priors, using simulated data. The density estimates are obtained by full data SMC and Subsampling SMC.  Subsampling MCMC fails to work on this example and hence its result is not included here.}
	\label{fig:comparekdeTN}
\end{figure}

Table \ref{tab:Experiment4} and Figure \ref{fig:comparekdeTN} summarize the results; they show that Subsampling SMC 
has difficulties exploring the skewed posteriors and gives inaccurate results when $m$ is too small. 
Our approach works poorly here because the posteriors for the first 5 individual effects are highly skewed,
and their skewness is caused by the truncated prior and the small number of observations. This causes the posterior to be concentrated at the tail of the log-density, where the control variates using a quadratic approximation are inaccurate. Therefore updating $\overline{\theta}$ by the posterior mean as specified in Algorithm \ref{alg:AIS-Algorithm-bigdata} does not produce good control variates, even though the log-density is well-behaved. Our likelihood estimate is inaccurate with high variance even when we use a slightly smaller subsample size compared to the previous section for estimating these skewed posteriors. We leave the development of more flexible control variates and 
the guidelines to choose an optimal subsample size, especially for complex posteriors, for future research.  Finally,
Subsampling SMC is not faster than full data SMC here because of the much larger $P$ and $R$ chosen by 
using the adaptive tuning method by \cite{buchholz2018adaptive}.




\begin{table}
	\centering
		\caption{Comparing the performance of the less accurate control variate (1st order) to the more accurate control variate (2nd order). The table shows the log of the estimate of the marginal likelihood (with standard errors in brackets), the CPU time, the number of annealing steps $P$ (tuned to maintain $ESS \approx 0.8M$) and the number of Markov moves $R$ (tuned as in \cite{buchholz2018adaptive}). The results are for the logistic regression model, estimated with the HIGGS dataset, using $M=280$ particles. The results are averaged over $10$ runs, which are used to compute the standard error of the estimator.}
	\fontsize{11}{11}\selectfont
	\begin{tabular}{*5c} \toprule
		& & &\\
		 & log marginal likelihood  & CPU time (hrs) & $P$ & $R$ \\ [5pt]\midrule
		& & \\
		\underline{\textbf{Logistic regression}} 
		 & & \\  [5pt]
		1st order &-7,013,461.07  &0.47  & 106 & 5 \\
		& \tiny{(0.46)} & &  \\[5pt]
		2nd order &  -7,013,460.90 &2.31  & 106 & 5 \\ 
		& \tiny{(0.32)} & & & \\ [5pt]
		
		\underline{\textbf{Truncated normal priors}} 
		 & & \\  [5pt]
		Full data SMC & -354,445.12   & 0.74 & 79 & 5 \\
		& \tiny{(0.26)}  & &  \\[5pt]
		Subsampling SMC & -354,437.34   & 1.13  & 141  & 36  \\ 
		& \tiny{(4.64)}  & & & \\ [5pt]
		\bottomrule
	\end{tabular}	
	\vspace{3mm}

	\label{tab:Experiment4}
\end{table}


\section{Application: Modeling firm bankruptcy nonlinearly\label{sec:RealApplication}}

The application of our method for model selection is now illustrated 
using a Swedish firm bankruptcy dataset containing $n = 4{,}748{,}089$ observations; the response variable is firm default and there are eight firm-specific and macroeconomic covariates, giving $9$ covariates, including an intercept. The data is treated as cross-sectional data and
the bank status is modeled by the logistic regression discussed in Section \ref{subsec:Experiment1}. 
A generalized additive model is also fitted to the data and is compared to a linear model; a similar prior $\boldsymbol{\theta} \sim \mathcal{N}(\mathbf{0}, 10^2 \mathbf{I}_{d})$ is used in both models. We compare the marginal posterior density estimates of Subsampling SMC against those of Subsampling MCMC \citep{Quiroz2018} as implemented by \cite{Dang2017} and find them nearly indistinguishable. We also compare both methods to the full data MCMC as in \cite{Dang2017}. However, it is unclear how to use Subsampling MCMC for model selection. Frequently used methods such as \cite{chib2001} are not useful for Subsampling MCMC since the (perturbed) likelihood cannot be evaluated; this is a major advantage of Subsampling SMC compared to Subsampling MCMC. 

We select between model $\mathcal{M}_1$ which is linear in the data on the logit scale and has $9$ coefficients, and
model $\mathcal{M}_2$ which is a semi-parametric additive model on the logit scale and uses B-splines as in \cite{Dang2017};  
model $\mathcal{M}_2$ is nonlinear in the data and has $81$ coefficients. 
Non-linear bankruptcy models for this dataset have previously been analyzed in \cite{quiroz2013dynamic} and \cite{giordani2014taking}. 
Given the marginal likelihood estimates,  the estimated Bayes Factor (BF) for the non-linear model $\mathcal{M}_2$ vs the linear model $\mathcal{M}_1$ is
\begin{equation}\label{eq:BayesFactor}
\wh {\mathrm{BF}}_{21} = \frac{\wh {\Pr}(\boldsymbol{y}|\mathcal{M}_2)}{\wh {\Pr}(\boldsymbol{y}|\mathcal{M}_1)};
\end{equation}
this is also the estimated ratio of posterior model probabilities when the prior model probabilities are equal. We use the 
strength of evidence guidelines in \citet[p. 438]{jeffreys1961theory} to choose between the models; Jeffreys 
considers  $10^{3/2} < \mathrm{BF}_{21} < 10^2$ as very strong evidence for model  $\mathcal{M}_2$ and $\mathrm{BF}_{21} > 10^2$ as decisive evidence. 

The number of blocks was set to $G=100$ with the subsample size set to $m=3{,}000$; for Subsampling MCMC these tuning parameters were
set as in \cite{Dang2017}. The estimates from the full data MCMC are considered as the ``gold standard'' when assessing the accuracy of the algorithms. This was achieved through an MCMC chain of $2{,}000$ post burnin MCMC samples, with the burnin  $=1{,}000$ iterations. The MCMC mixed well and we believe that the iterates represent the posterior adequately .

Table~\ref{tab:marginal_likelihood_and_BF} reports the estimated log of the marginal likelihood for both models and the corresponding Bayes factors obtained by Subsampling SMC. The table shows decisively that the non-linear model is superior. We again stress that producing marginal likelihood estimates is very convenient by SMC, whereas it is currently not possible with Subsampling MCMC.

\begin{figure}[H]
\centering{}\includegraphics[width=0.8\linewidth]{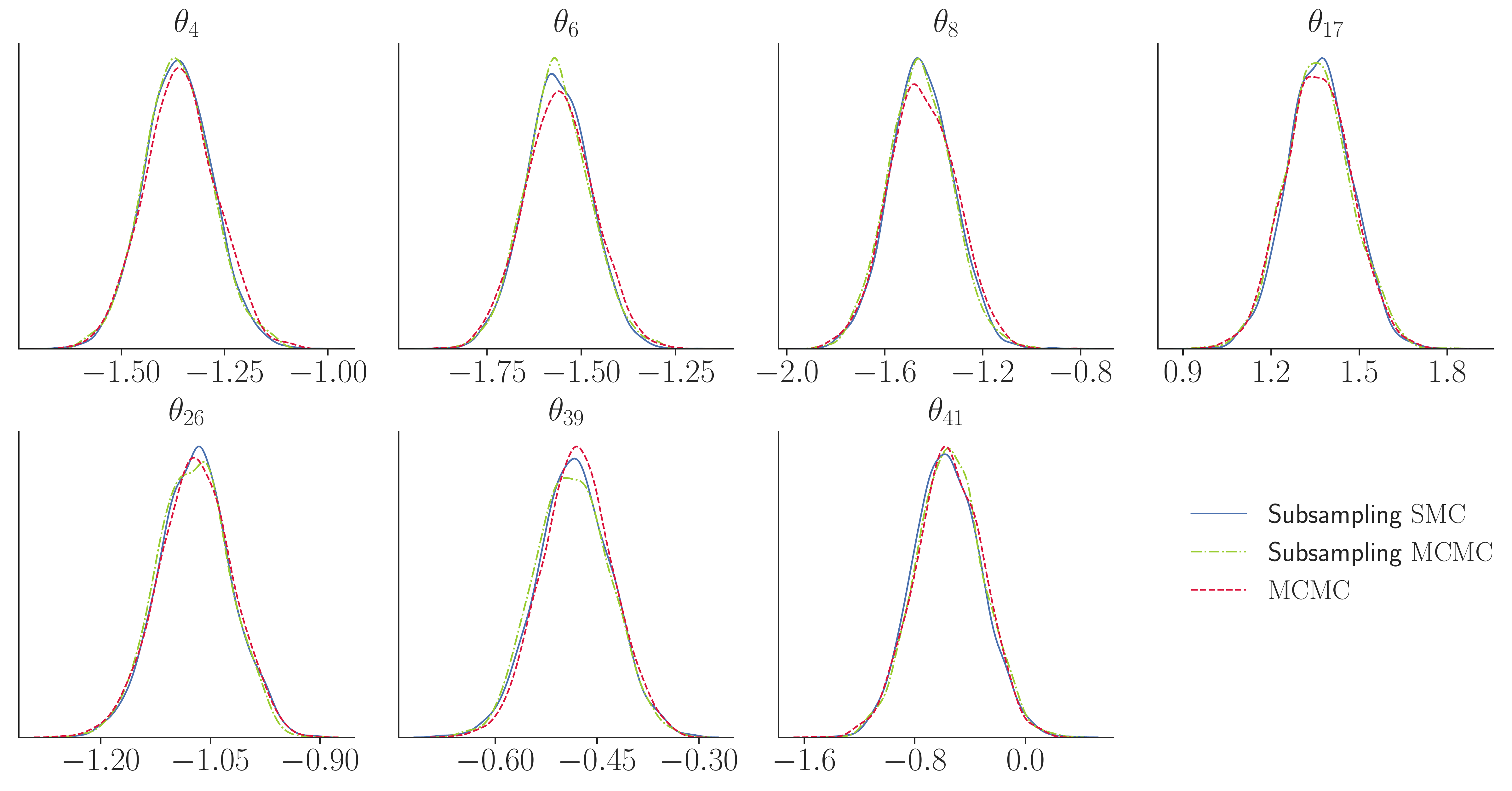}
\caption{Kernel density estimates of a subset of the marginal posterior densities of $\boldsymbol{\theta}$
for the logistic model $\mathcal{M}_2$ for the bankruptcy dataset. The density estimates are obtained by MCMC, Subsampling MCMC and Subsampling SMC. MCMC represents the ground truth.
\label{fig:The-Kernel-Density thetacol6_bankruptcy}}
\end{figure}

\begin{figure}[H]
\centering{}\includegraphics[width=1\linewidth]{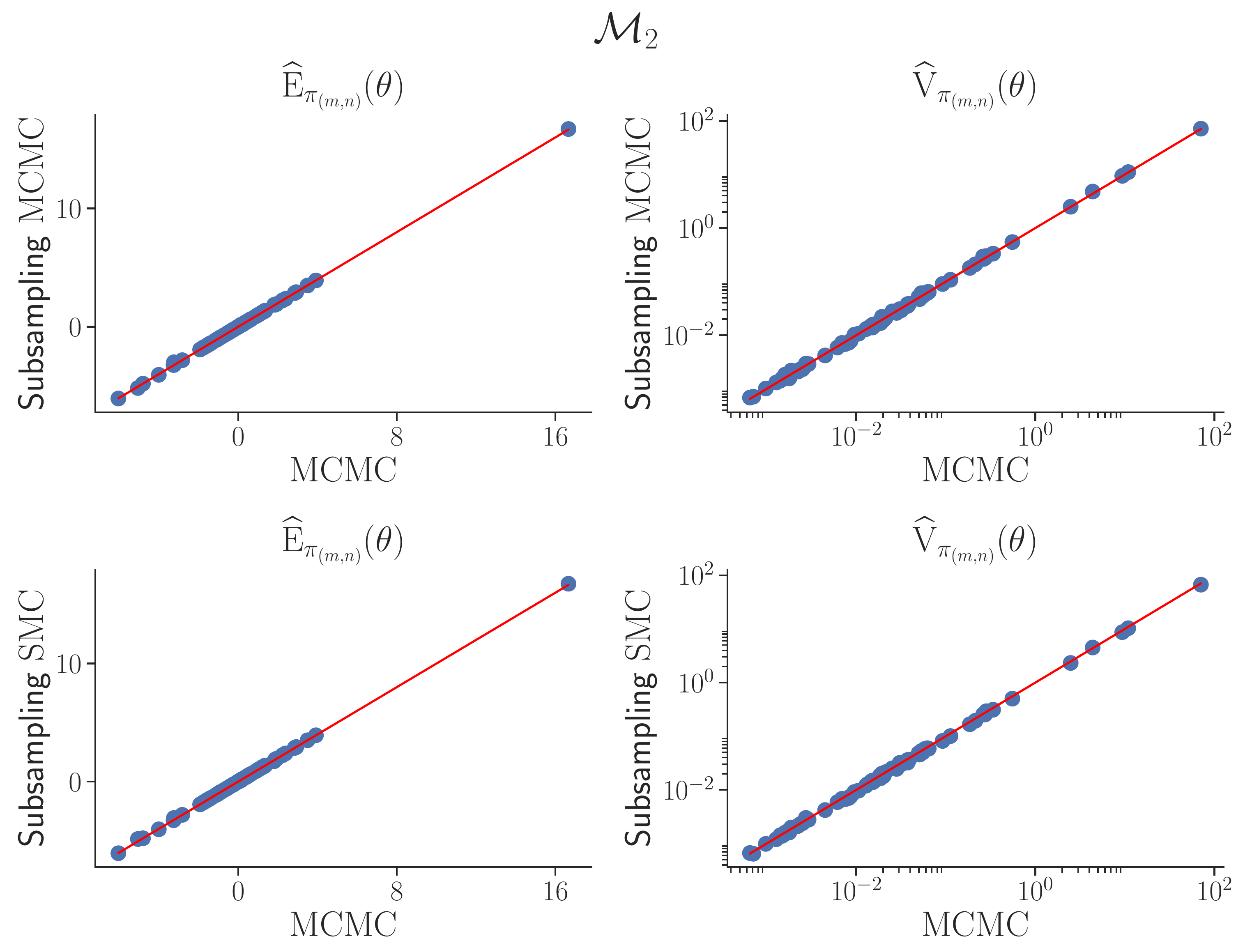}
\caption{Estimates of marginal posterior means (left panel) and posterior variances (right panel) of $\boldsymbol{\theta}$
for the logistic model $\mathcal{M}_2$ for the bankruptcy dataset. The estimates are obtained by Subsampling MCMC and Subsampling SMC and plotted as dots, together with a 45 degree line which corresponds to estimates that are in perfect agreement.
\label{fig:Mean_var_bankruptcy}}
\end{figure}

\begin{table}
\centering \caption{Log of the estimates of the marginal likelihoods and Bayes factors $\mathrm{BF}_{21}$
in \eqref{eq:BayesFactor} for selecting between $\mathcal{M}_{1}$ and $\mathcal{M}_{2}$. The estimates of the Standard Errors (SE) are in brackets. The SE is computed using the $10$ independent parallel runs. The prior probabilities are $\Pr(\mathcal{M}_{1})=\Pr(\mathcal{M}_{2})=1/2$.}
\vspace*{2mm}
\begin{tabular}{lrrrrrr}
\toprule
 & $\log\widehat{p}(\boldsymbol{y}|\mathcal{M}_{1})$ &  & $\log\widehat{p}(\boldsymbol{y}|\mathcal{M}_{2})$ &  & $\mathrm{\wh BF}_{21}$ & \tabularnewline
\cmidrule{2-2} \cmidrule{4-4} \cmidrule{6-6}
Bankruptcy  & $\underset{(0.21)}{-208{,}517.79}$ &  & $\underset{(6.57)}{-200{,}215.13}$ &  & $\underset{~}{\exp(8{,}302.66)}$  & \tabularnewline
\bottomrule
\end{tabular}\label{tab:marginal_likelihood_and_BF}
\end{table}
Figures \ref{fig:The-Kernel-Density thetacol6_bankruptcy} shows the kernel density estimates of the marginal posterior of selected parameters of the non-linear model for the bankruptcy dataset. It is evident that both Subsampling SMC and Subsampling MCMC are very accurate and we have confirmed the accuracy of the kernel density estimates for all the parameters, which we do not show here to save space. Instead, Figure \ref{fig:Mean_var_bankruptcy} shows the estimated marginal posterior expectations and posterior variances by the two algorithms for all the parameters in the non-linear models. This confirms the accuracy of the estimates of each parameter. We have also confirmed that the kernel density estimates and the estimated marginal posterior expectations and posterior variances are accurate for the linear model (not shown here). 

\begin{figure}[H]
\centering{}\includegraphics[width=\linewidth]{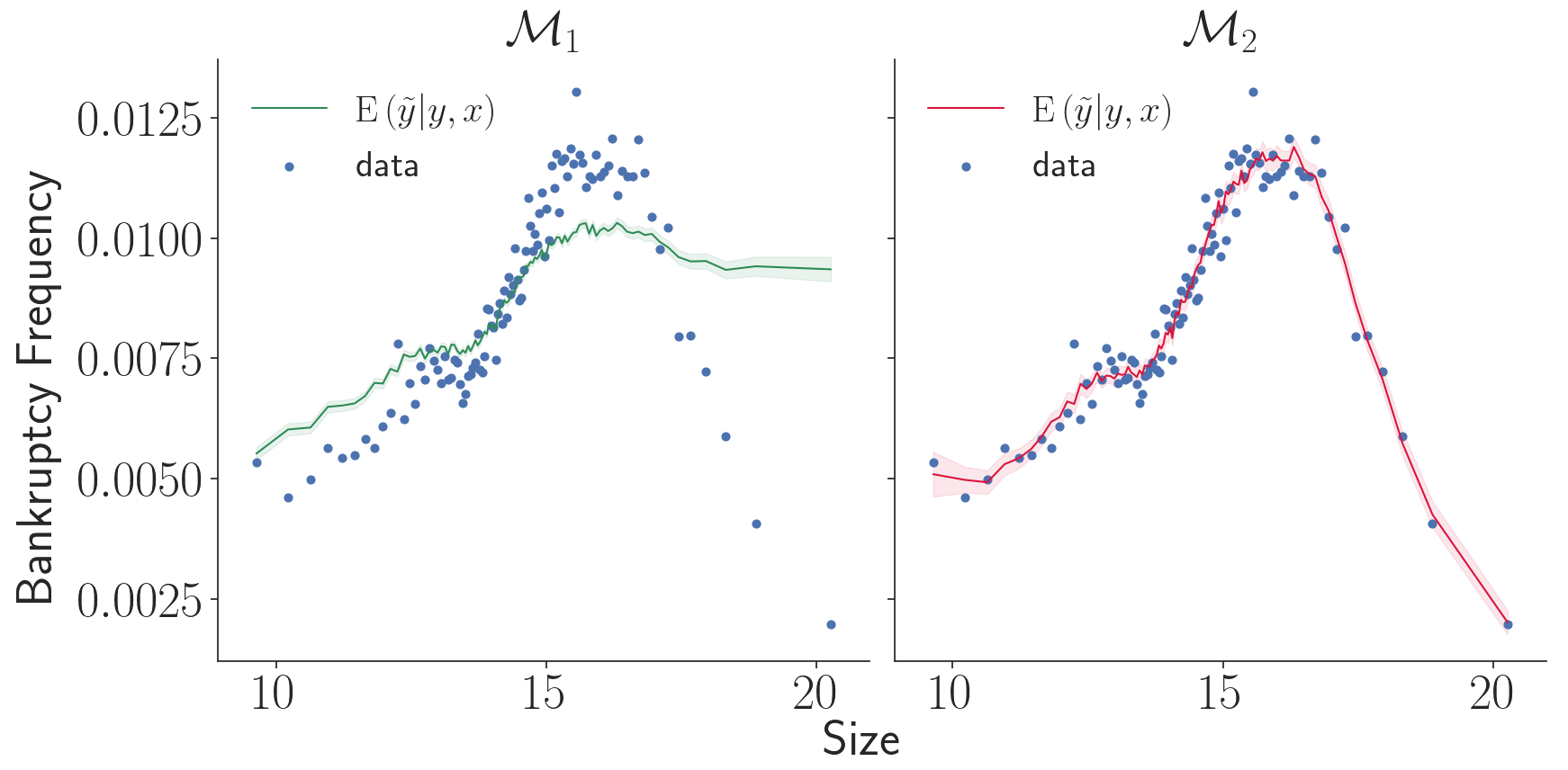}
\caption{Realized and estimated bankruptcy probabilities. The figure shows the results with respect to the size variable (logarithm of deflated sales) for $\mathcal{M}_1$  (left panel) and $\mathcal{M}_2$  (right panel). The data are divided into $100$ equally sized groups based on the size variable. For each group, the empirical estimate of the bankruptcy probability is the fraction of bankrupt firms. These empirical estimates are represented as dots, where the corresponding $x$-value (size) has been set to the mean within the group. The model estimates for each of the $100$ groups are obtained by, for each posterior sample $\boldsymbol{\theta}$, averaging the posterior predictive $\Pr\left(\widetilde{y}_k=1|\boldsymbol{y}, x_k\right)$ for all observations $k$ in a group, and subsequently computing the posterior predictive mean $\E\left(\widetilde{y}_k=1|\boldsymbol{y}, x_k\right)$ (solid line) and $90$\% prediction interval (quantiles $5\mbox{-}95$, shaded region). \label{fig:Non-linear_predictions_M1_vs_M2}}
\end{figure}

Figure \ref{fig:Non-linear_predictions_M1_vs_M2} shows that the relationship between the probability of bankruptcy and the covariate Size is not a logistic function (inverse-logit) of the covariate and that the nonlinear model fits the data much better than the linear logistic model.

\section{Conclusions\label{sec:Conclusions}}
A simple and effective approach is proposed to speed up sequential Monte Carlo for static Bayesian models using data subsampling. Its key ingredients  are an efficient annealed likelihood estimator and an effective Markov kernel move step based on Hamiltonian Monte Carlo to boost particle diversity. This kernel is computationally expensive for large datasets and data subsampling is crucial to obtain a feasible approach. We argue that the subsampling approach is also very convenient for managing computer memory when implementing SMC using parallel computing, because it avoids the need for each worker to store the full dataset. We demonstrate that the method performs efficiently and accurately for four generalized linear models and a generalized additive model. Moreover, it allows Bayesian model selection through accurate estimates of the marginal likelihood, which is a major advantage compared to Subsampling MCMC. We also illustrate that the limitation of our method is that its performance depends on good control variates, which can be challenging to construct in certain models. An anonymous reviewer suggested we may use the SMC particles to construct a surrogate function to use as control variate in more complex models. How to do this in a computationally efficient way is an open question, and we leave this extension for future research


\section*{Acknowledgements}
We thank the Associate Editor and two reviewers for helping to improve both the content and the presentation of the article. 
Khue-Dung Dang, David Gunawan, Matias Quiroz and Robert Kohn were partially supported by
Australian Research Council Center of Excellence grant CE140100049.

\bibliographystyle{apalike}
\bibliography{references_v1,pfmcmc}

\appendix

\end{document}